\documentclass[10pt,twocolumn,letterpaper]{article}

\usepackage{cvpr} 
\usepackage{graphicx}
\usepackage{soul,color}
\usepackage{amssymb}
\usepackage{booktabs}
\usepackage{caption}
\usepackage{subcaption}
\usepackage{float}

\usepackage[pagebackref,breaklinks,colorlinks]{hyperref}

\usepackage[capitalize]{cleveref}
\crefname{section}{Sec.}{Secs.}
\Crefname{section}{Section}{Sections}
\Crefname{table}{Table}{Tables}
\crefname{table}{Tab.}{Tabs.}

\def\cvprPaperID{*****} 
\def\confName{CVPR}
\def\confYear{2022}

\title{AI-Enhanced Data Processing and Discovery Crowd Sourcing for Meteor Shower Mapping}

\begin{document}





\author{Siddha Ganju\thanks{\noindent\textsuperscript{}Email IDs  of corresponding authors: \\ sganju@nvidia.com, \\ amartya.hatua@fmr.com}\\
Nvidia\\ 
SpaceML\\
Frontier Development Lab
\and
Amartya Hatua{\small$^*$}\\
SpaceML\\
Fidelity Investments\\
\and
Peter Jenniskens\\
SETI Institute\\
\and
Sahyadri Krishna\\
SpaceML\\
IISER, Tirupati\\
\and
Chicheng Ren\\
Square\\
SpaceML\\
\and
Surya Ambardar\\
SpaceML\\
University of Virginia\\
}

\maketitle

\begin{abstract}
The Cameras for Allsky Meteor Surveillance (CAMS) project, funded by NASA starting in 2010, aims to map our meteor showers by triangulating meteor trajectories detected in low-light video cameras from multiple locations across 16 countries in both the northern and southern hemispheres. Its mission is to validate, discover, and predict the upcoming returns of meteor showers. Our research aimed to streamline the data processing by implementing an automated cloud-based AI-enabled pipeline and improve the data visualization to improve the rate of discoveries by involving the public in monitoring the meteor detections. This article describes the process of automating the data ingestion, processing, and insight generation using an interpretable Active Learning and AI pipeline.  This work also describes the development of an interactive web portal (the NASA Meteor Shower portal) to facilitate the visualization of meteor radiant maps. To date, CAMS has discovered over 200 new meteor showers and has validated dozens of previously reported showers.
\end{abstract}

\section{Introduction}
\label{sec:intro}

Since the beginning of civilization, the night sky has intrigued mankind with its mysteries, provoking the most existential questions of all - ``Are we alone? How did we get here?" Meteors and other rocky objects from space may hold the answer to these questions. Speculation exists that meteors and related rocky objects may have brought water and amino acids, the building blocks of life, to Earth. Learning about meteors and space debris involves collecting them when they fall to Earth, tracking them in space, and tracing their origins. At the SETI Institute, CAMS was founded to map meteors to their parent bodies in space. 

First CAMS sites was set up in the San Francisco Bay area on October 10, 2010, with low-cost cameras stationed at the Lick Observatory, Fremont Observatory, and the rooftop of a Mountain View home. Over time, these cameras have been installed for night sky observation in 600 sites across the globe, with each site capturing an average of 500 moving light detections every night from 16 cameras. For the last 13 years, these cameras have been recording observations of the night sky, and these observations have been classified manually until the automated cloud-based AI-enabled pipeline is implemented. Each camera collects 10 GB of data on an average night, making the manual process extremely time-consuming. In addition, each camera is expected to record thousands of observations on nights when meteor showers occur.

Over the years, the number of observation stations increased, so the amount of data has also increased. Now the manual effort required for domain scientists to retrieve and analyze this data increased exponentially. We sought to harness the power of AI to mitigate the time consuming manual review process and accelerate new meteor discovery. Inspired by advancements in machine learning techniques,  the CAMS data network is automated. The automation eliminates the need to travel to each location and download camera information. It also allows cameras to function independently, with minimum manual intervention. Additionally, a machine learning pipeline is trained to replicate scientists' thought processes and classify night sky activities as meteors of interest or other streaks of light in the night sky like planes or satellites. In conjunction, an active learning pipeline points out exciting activities in the night sky that may not be recognized as meteors. In order to increase the transparency and interpretability of the insights offered by the ML pipeline, we present them in an intelligent interactive webportal, NASA CAMS Meteor Shower portal \cite{setiNASACAMS}. It allows users to view the activity of the night sky at the click of a button. The CAMS portal brings visibility to rare, sporadic, and known meteor shower activity from the previous night, from anywhere in the world, onto an internet browser. Perhaps more interestingly, it facilitates the visualization of and discovery of patterns in data, aiding in discovery of meteors' parent bodies and other phenomena. 


There are several projects and networks that aim to track meteor and fireball activity across the globe. One such network is the Desert Fireball Network (DFN) \cite{dfs} in Australia. The DFN aims to locate fallen fireballs and employs a pipeline similar to CAMS. The DFN uses high-resolution images as opposed to the low-resolution security cam videos used by CAMS. The DFN dissects images into smaller sections called tiles and feeds these tiles for classification as fireball or non-fireball by a neural network. The DFN reported an 82$\%$ detection rate for small fireballs (a single tile), but 100$\%$ for large fireballs (stretched over multiple tiles). The DFN also reported a higher false-positive rate for small fireballs compared to large fireballs. Since large fireballs are more important in the context of fireball retrieval, the reported detection rates satisfied the project’s goals. Another similar project is the Canary Islands Long-Baseline Observatory (CILBO) \cite{gonzalez2018systematic} operated by the European Space Agency (ESA). CILBO uses a meteor detection pipeline that, unlike the CAMS meteor pipeline, uses spectroscopy gratings to obtain spectra in addition to images and videos. The CILBO pipeline performs three tasks: A) Selects meteors with spectra, B) Extracts spectra for meteors within the wavelength range of 400 to 800 nm, and C) Analyses spectra to decipher meteor composition. The Fireball Recovery and InterPlanetary Observation Network (FRIPON) \cite{bland2012australian} has objectives as the CAMS project. Both estimate the trajectories of meteor showers, and both aim to characterize meteors and their parent bodies. The FRIPON network, with 150 cameras and 25 radio receivers set up in Europe and Canada, uses similar software as CAMS. The FRIPON pipeline has routines to calculate meteor trajectory factoring in atmospheric entry disintegration, which are tested using synthetically generated meteors.

Unlike the above networks, the CAMS network feeds raw data directly to the CAMS pipeline. We work on unprocessed raw data rather than image modality, as raw data is lighter and thus relatively inexpensive to transfer. Additionally, unlike several of the above meteor or fireball orbit surveys which are local to a country or continent, CAMS is a global network, with stations in 12 countries giving 24x7 coverage of the entire night sky. Scaling CAMS or setting up a new station is relatively easy and involves two main steps: 
1) setting up the hardware (camera and a computer station), 2) downloading and running prebaked scripts that control the camera and transfer data. 

In this paper, we describe an AI-based meteor identification process which helps researchers to scale the meteor discovery process immensely. The process starts with data collection and transfer, followed by data preprocessing and augmentation techniques. Once processed and augmented, the data is used for meteor classification using a deep-learning-based bi-directional LSTM model. The classifier has performed efficiently and achieved high accuracy, precision, recall, and F1 scores.  To better understand patterns embedded in the data and provide a visual interpretation of the classifier, an attention-based deep learning model has been used. Post-classification, CAMS backend system helps with data querying, API creation, and related tasks to develop an asynchronous job handling system. This backend system helps develop a real-time meteor information system available on the NASA CAMS Meteor Shower Portal. In the following sections, we describe each of these steps, provide information about 10 newly discovered meteor showers, and describe some observations of rear meteor showers.


\section{Data Collection}
\label{sec:data_collection}

With an average of 24,000 observations per night, per location, manual verification by domain scientists quickly bec intractable and difficult to scale. On nights with putative showers such as Perseids or Geminids, increased meteor shower rates result in an astronomically higher number of observations, taking weeks to sift through manually. Seamless data transfer and processing was thus necessary to scale up the CAMS system to handle more incoming data and set up new stations globally with minimal effort. Previously, due to manual verification and transfer, the data processing pipeline built up a backlog of 3-4 months. We parallelized the data processing pipeline, resulting in an average performance improvement of 50$\%$. Now, data processing is real-time with global meteor information available on the NASA CAMS Meteor Shower Portal the next morning. A detailed description of the web portal is presented in the ``CAMS Portal" section.

\section{Data Processing for AI Model}
\label{sec:data_preprocessing}

A key component of the CAMS project is an AI-based meteor identification model. The data used to train this model was combined from three triangulation stations into multiple sequences. Data domain knowledge and initial experiments showed that five unique characteristics such as the latitude, and longitude of the meteor; calculated object height and speed; and brightness in terms of visual magnitude are most informative in distinguishing trends in the movements of objects. Therefore, these features are used as inputs for training and testing meteor deep learning based classification model.

\begin{figure*}
     \centering
         \centering
         \includegraphics[width=0.9\textwidth]{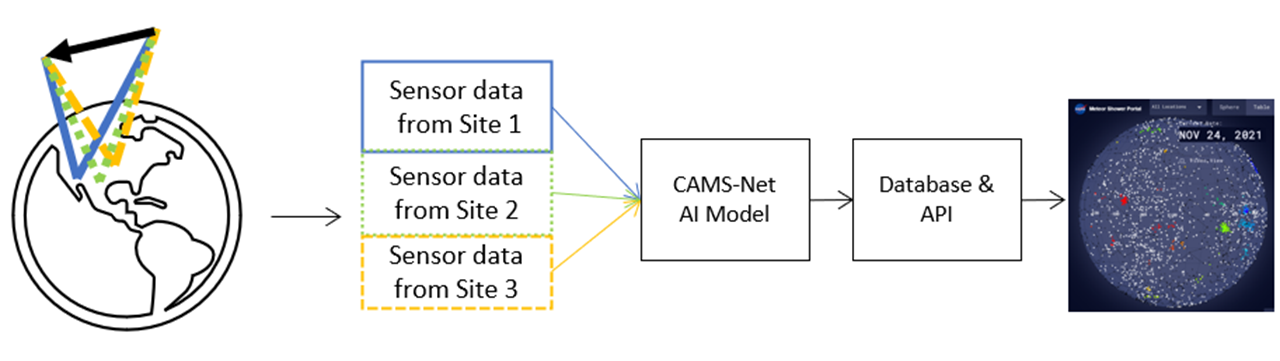}
         \caption{CAMS Architecture}
         \label{fig:cams_net}
\end{figure*}

The trained model generates inferences on raw camera data rather than images or videos. Sensor data occasionally contains noise or missing data, which is replaced by the median of the rest of the data. The features of training and testing dataset are normalized to zero mean, and unit variance. We have released the code that uses these standard guidelines for processing the decade-long meteor dataset.

\subsection{Data Augmentation Techniques}
\label{sec:data_augmentation}

CAMS possesses a year's worth of data manually annotated by scientists for AI-based model training, with 4,510 examples of true meteors and 29,600 examples of non-meteors. This is similar to the observed distribution of night sky events - many non-meteors and only a few true meteors are visible on any given night. Additionally, data exploration revealed that most meteors are captured by cameras for a relatively short duration when compared to other objects like the lights of planes and satellites. Providing the differences in durations for model training could result in improved performance.

Before the data is ready for modeling, it undergoes various preprocessing stages. To better analyze the data, we have divided each data point into smaller units or frames. Each frame corresponds to a specific time period and contains observation data. To maintain consistency in the data, we use a windowing technique that creates data points of a fixed number of frames. This is necessary because each observation has varying numbers of frames. Padding is used when there are insufficient frames available. For longer observations that need to be divided into smaller ones, this guarantees a consistent length. In order to standardize training instances and maximize the number of examples, it is imperative to break down longer meteor instances into multiple instances. This approach ensures a more streamlined and powerful training process. Data with different window sizes (15, 20, 30, 35, 40, 45, 50, 55, 60) was tested, and we observed that a window size of 45 gives the best result. Results from training baseline models for different lengths are shown in Table \ref{tab:result_lstm}.

\begin{table*}[]
\begin{tabular}{|c|c|c|c|c|c|c|c|c|c|c|c|}
\hline
Window size & Train & + in Train & Val   & +in Val & TP   & TN    & FP   & FN  & Precision & Recall & F1 Score \\ \hline
15          & 92014 & 66674      & 10224 & 7379    & 7379 & 10224 & 2845 & 0   & 0.72      & 1.0    & 0.83     \\ \hline
20          & 65337 & 47866      & 7260  & 5375    & 5121 & 7006  & 329  & 254 & 0.93      & 0.95   & 0.93     \\ \hline
25          & 48234 & 36028      & 5360  & 4011    & 3856 & 5205  & 350  & 155 & 0.91      & 0.96   & 0.93     \\ \hline
30          & 37765 & 28541      & 4197  & 3134    & 2973 & 4036  & 251  & 161 & 0.92      & 0.94   & 0.92     \\ \hline
35          & 29313 & 22668      & 3258  & 2526    & 2461 & 3193  & 205  & 65  & 0.92      & 0.97   & 0.94     \\ \hline
40          & 23928 & 18788      & 2659  & 2102    & 2034 & 2591  & 121  & 68  & 0.94      & 0.96   & 0.95     \\ \hline
45          & 19152 & 15420      & 2128  & 1725    & 1650 & 2053  & 199  & 75  & 0.89      & 0.95   & 0.91     \\ \hline
50          & 16110 & 13170      & 1790  & 1482    & 1417 & 1725  & 94   & 65  & 0.93      & 0.95   & 0.93     \\ \hline
55          & 13103 & 11043      & 1456  & 1205    & 1205 & 1456  & 251  & 0   & 0.82      & 1.0    & 0.90     \\ \hline
60          & 11158 & 11043      & 1456  & 1205    & 1205 & 1456  & 251  & 0   & 0.82      & 1.0    & 0.90     \\ \hline
65          & 9258  & 8050       & 1029  & 902     & 902  & 1029  & 127  & 0   & 0.87      & 1.0    & 0.93     \\ \hline
\end{tabular}
\caption{Baseline model result}
\label{tab:result_lstm}

\end{table*}

However, this windowing technique suffers information loss either when the generated smaller-length meteors or the original data from three cameras are shorter than the length threshold. We also note that the windowing technique adds ambiguity as it considers combined data rather than data from each camera independently. Three camera sites per observation location may observe a streak of light for different durations and maintaining temporally consistent input sequences may impact model performance. To reduce information loss and ambiguity, a Multi-Camera Context (MCC) data processing technique was used which first splits data from each camera individually into segments (rather than combining data from all cameras in windowing), and then combines them together to generate a single observation.

\begin{table*}[]
\centering
\begin{tabular}{|l|c|c|c|c|}
\hline
Model Architecture             & \multicolumn{1}{l|}{Precision} & \multicolumn{1}{l|}{Recall} & \multicolumn{1}{l|}{F1 Score} & \multicolumn{1}{l|}{Best window size} \\ \hline
Baseline                       & 0.9400                         & 0.9600                      & 0.9499                        & 40                                    \\ \hline
LSTM-1 & 0.9640                         & 0.9622                      & 0.9631                        & 65                                    \\ \hline
LSTM-2 & 0.9527                         & 0.9748                      & 0.9636                        & 35                                    \\ \hline
LSTM-3 & 0.9652                         & 0.9645                      & 0.9649                        & 30                                    \\ \hline
GRU-1                          & 0.9585                         & 0.9665                      & 0.9625                        & 40                                    \\ \hline
GRU-2                          & 0.9632                         & 0.9813                      & 0.9721                        & 65                                    \\ \hline
GRU-3                          & 0.9692                         & 0.9757                      & 0.9724                        & 30                                    \\ \hline
CNN-LSTM-1                     & 0.9604                         & 0.9654                      & 0.9629                        & 40                                    \\ \hline
CNN-LSTM-2                     & 0.9675                         & 0.9683                      & 0.9679                        & 55                                    \\ \hline
CNN-LSTM-3                     & 0.9574                         & 0.9643                      & 0.9608                        & 35                                    \\ \hline
CNN-GRU-1                      & 0.9625                         & 0.9675                      & 0.9650                        & 40                                    \\ \hline
CNN-GRU-2                      & 0.9633                         & 0.9737                      & 0.9685                        & 45                                    \\ \hline
CNN-GRU-3                      & 0.9734                         & 0.9702                      & 0.9718                        & 65                                    \\ \hline
LeakyPostReg                   & 0.9500                         & 0.9630                       & 0.9565                        & 45                                    \\ \hline
P-CNN-GRU-3                    & 0.9470                         & 0.9874                      & 0.9668                        & 45                                    \\ \hline
P-BiLSTM                       & 0.9691                         & 0.9777                      & 0.9734                        & 45                                    \\ \hline
P-CNN-BiLSTM-1                 & 0.9640                         & 0.9832                      & 0.9735                        & 45                                    \\ \hline
P-CNN-BiLSTM-2                 & 0.9638                         & 0.9827                      & 0.9732                        & 45                                    \\ \hline
\end{tabular}
\caption{Comparison between performaces of differnt models}
\label{tab:result_comparison_model}
\end{table*}

\subsection{Training and Inference of CAMS-Net Meteor Detector}
\label{sec:training_inference}

To determine the optimal predictive model, we conducted experiments with four distinct neural network architectures: LSTM, CNN-LSTM, GRU, and CNN-GRU. The baseline model architecture \ref{model:baseline} employs two stacked bidirectional LSTM (BiLSTM) layers and four dense layers to map the combined hidden dimensions of the LSTM layers to the output tag space, which is binary and one-dimensional. Each layer has a RELU activation function, except for the last Dense layer, which has sigmoid activation suited for binary classification. The loss function used is binary cross entropy with an Adam optimizer. The model converges on an Nvidia TitanX GPU in a few hours. The baseline model is trained and tested using data with all possible window sizes (15, 20, 30, 35, 40, 45, 50, 55, 60). In Table \ref{tab:result_lstm} performance of the baseline model is presented.  The first column represents the Window size for each data. Train and +in Train represent the number of training data and data with positive class; while Val and +in Val represent the number of validation data and data with positive class. The rest of the columns are representing, True Positive (TP), True Negative (TN), False Positive (FP), False Negative (FN), Precision, Recall, and F1 Score. The results show that the  baseline model perfoms best for data with window size 45. 

Beside the baseline model, the similar experiment have been performed with eighteen models such as LeakyPostReg, P-CNN-GRU-3, P-BiLSTM, P-CNN-BiLSTM-1, P-CNN-BiLSTM-2,  LSTM-1, LSTM-2, LSTM-3, GRU-1, GRU-2, GRU-3, CNN-BiLSTM-1, CNN-BiLSTM-2, CNN-BiLSTM-3, CNN-GRU-1, CNN-GRU-2, CNN-GRU-3. A detailed description of the these models are provides in Appendix A \ref{appendix_a}. Comparing the performances of all the models, we found P-CNN-BiLSTM-2 model prodcues best results for classification. For this model the window size is 45.  

\section{Active Learning pipeline}
\label{sec:active_learning}
Supervised classification models require large amounts of labeled dataset for training. Labeling a dataset, is conventionally a manual task requiring extended effort. To address this issue, we leveraged one year's worth of labeled meteor activity data from CAMS to provide weak supervision for over a decade of collected data, drastically reducing the amount of manual annotation necessary and expanding the available labeled meteor training data. We utilized one year's labeled data to train a high-confidence LSTM meteor classifier to generate low-confidence labels for the remaining decade’s worth of meteor data. Our classifier yields confidence levels for each prediction. When the confidence lies above a statistically significant threshold (95$\%$ precision and 95$\%$ recall), predicted labels can be treated as weak supervision for future training runs. The remaining predictions which are below the threshold (i.e. weak confidence) can be manually annotated before being added to the training set for future training. This adds a human-in-the-loop component with expert-guided testing to verify all the low-confidence meteors. Using a high threshold minimizes label noise and ensures instances are correctly labeled, while dramatically reducing the amount of data that needs to be manually annotated by domain scientists. We also used the active learning pipeline to automate insight generation and learn about unique artifacts in different geographic regions where the CAMS cameras are set up. For example, camera boxes in the UAE are often covered with sand, and footage from cameras in Australia and New Zealand contains fireflies. Such artifacts demand retraining to distinguish them from meteors. 

\begin{figure*}
  \begin{subfigure}[b]{0.5\linewidth}
    \centering
    \includegraphics[width=0.9\linewidth, height = 4cm]{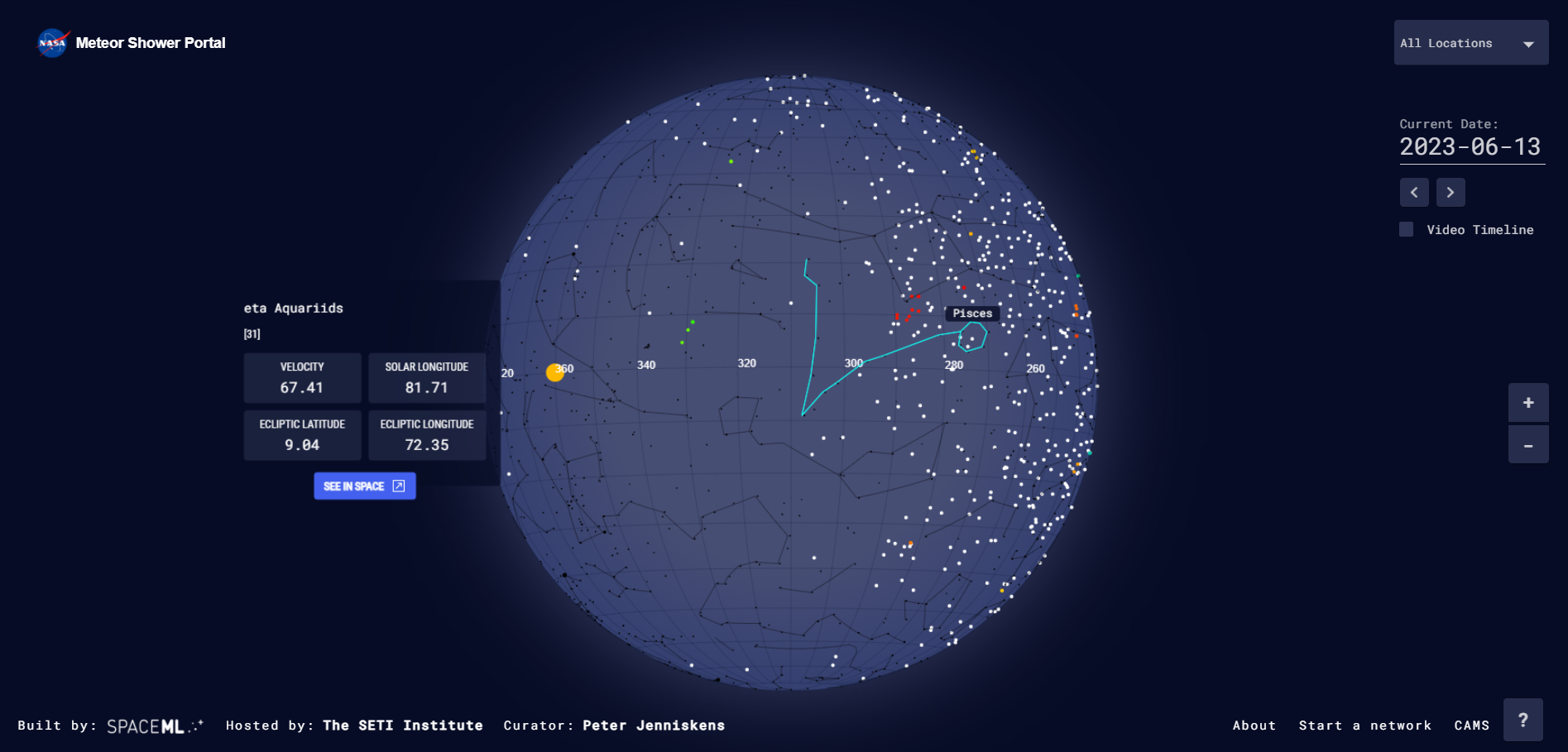} 
    \caption{Constellation view Pisces and features of eta Aquariids} 
    \label{web_portal:a} 
    \vspace{4ex}
  \end{subfigure}
  \begin{subfigure}[b]{0.5\linewidth}
    \centering
    \includegraphics[width=0.9\linewidth, height = 4cm]{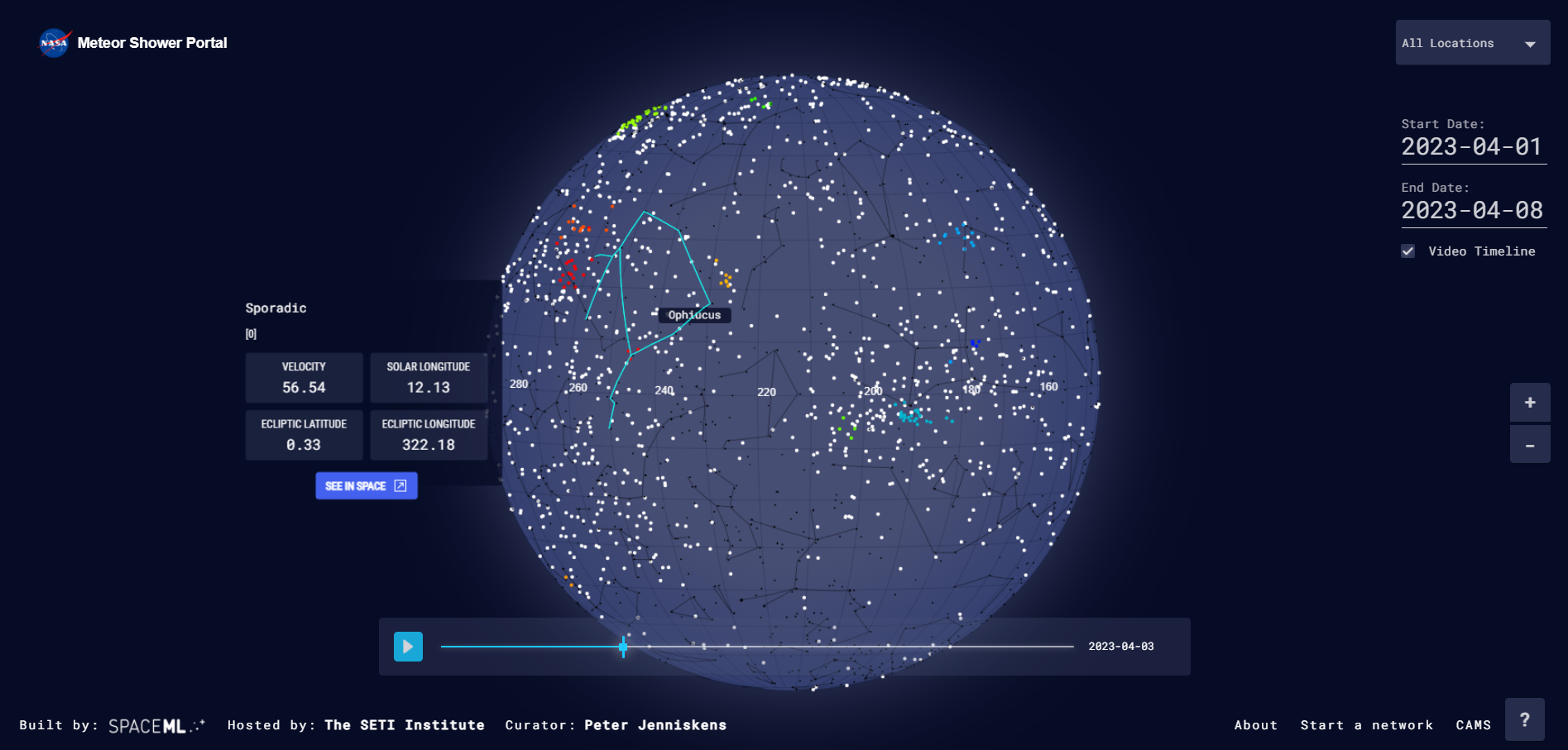} 
    \caption{Constellation view Ophiucus and video view } 
    \label{web_portal:b} 
    \vspace{4ex}
  \end{subfigure} 
  \begin{subfigure}[b]{0.5\linewidth}
    \centering
    \includegraphics[width=0.9\linewidth, height = 4.2cm]{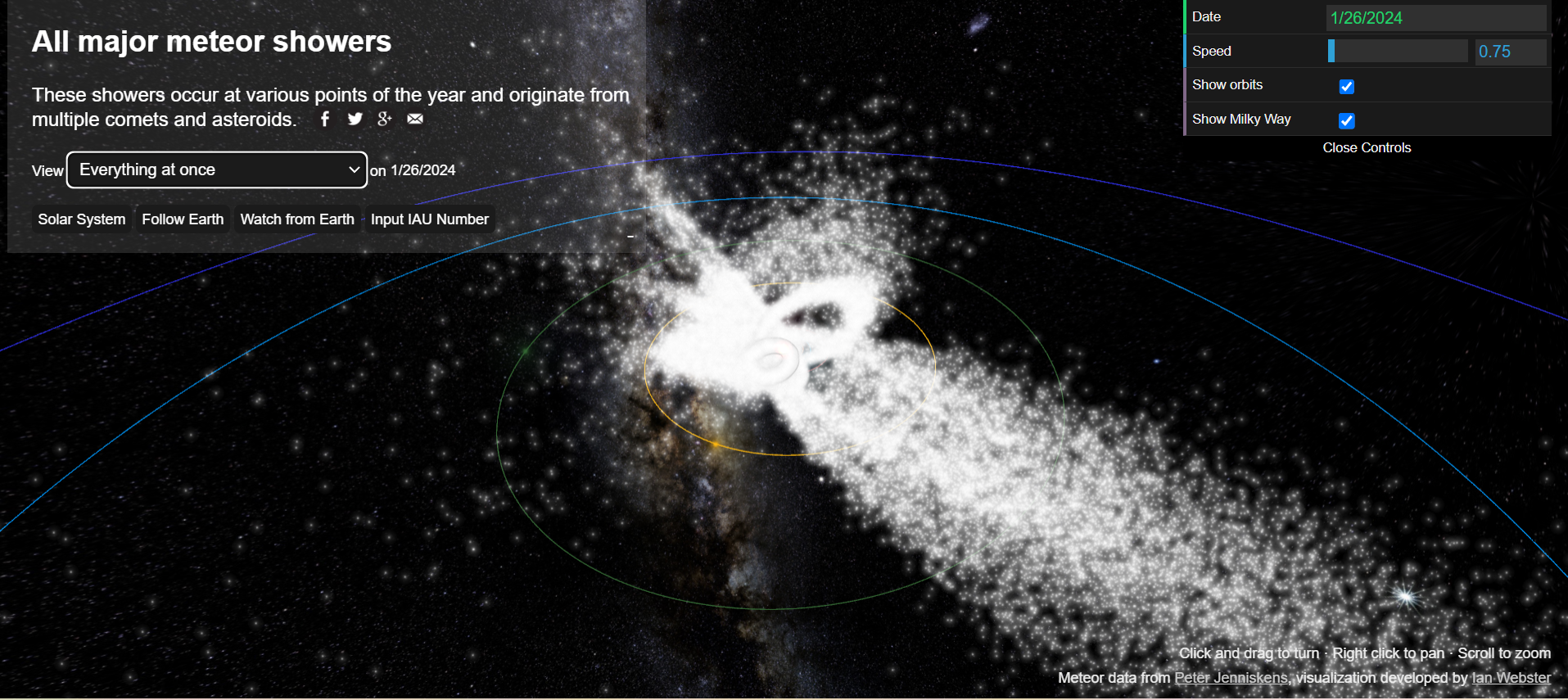} 
    \caption{Space view of all major meteor showers at once} 
    \label{web_portal:c} 
  \end{subfigure}
  \begin{subfigure}[b]{0.5\linewidth}
    \centering
    \includegraphics[width=0.9\linewidth, height = 4.2cm]{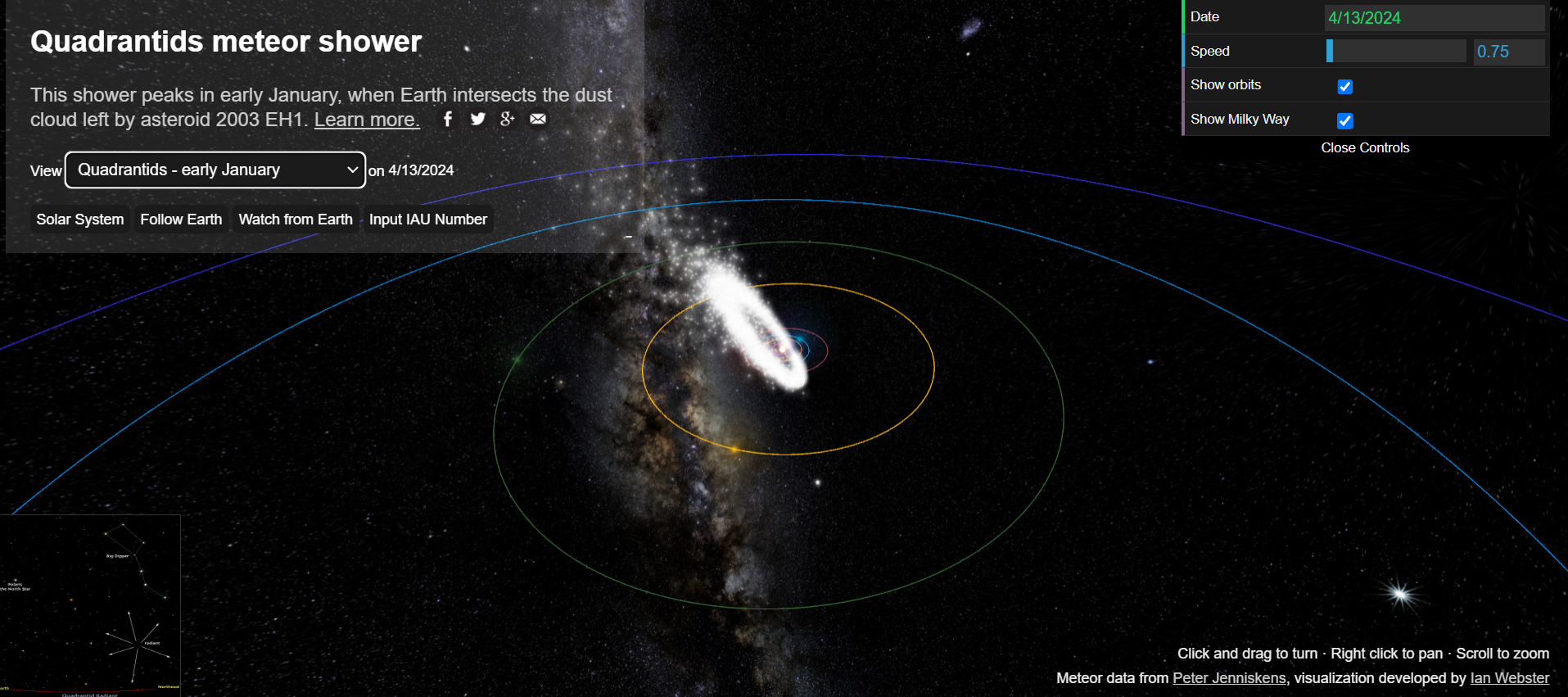} 
    \caption{Space view of Quadrantids} 
    \label{web_portal:d} 
  \end{subfigure} 
  \caption{Illustration of Web portal \cite{githubGitHubSidgancamspapervids}}
  \label{fig7} 
\end{figure*}

\section{Web portal view}
\label{sec:data_ingestion}

Once data is processed to identify meteors and non-meteors, the data can be viewed by scientists as global, nightly activity, or as a yearly summary. To construct complex queries such as ``show all meteors for Geminids 2021", a CAMS Backend System is developed which establishes (1) a database system for fast querying, (2) an Application Programming Interface (API) for rich querying, and (3) an asynchronous job handling mechanism. When data is consumed in the backend system, each data entry is stored as a row in our designated database tables. With proper indexing set up on the table, the target dataset can be efficiently queried with a simple database query. Additionally, the database grants us the possibility to conduct more complicated database-level data aggregation in the future. With the web application setup, APIs are exposed to interact with meteors, stars, constellations related data, and other system endpoints, allowing the website to be fast and lightweight. Finally, an asynchronous job handling system allows us to incorporate long-running jobs such as timeline generation in the backend. The backend system additionally sends daily alerts via Slack, detailing statistics such as the number of observations processed, the number of meteors detected in any event that requires visual inspection by the scientists, etc. to provide visibility across all processes.

\section{CAMS Portal}
\label{sec:cams_portal}
In addition to identifying meteors in the night sky, we introduce an interactive portal \cite{setiNASACAMS}.  It publishes the meteor detection information till the night before the portal is visited. This web portal also includes a state-of-the-art user-centered design and several new features powered by an upgraded data ingestion pipeline. The NASA CAMS Meteor Shower Portal follows the heliocentric ecliptic coordinate system where the coordinates of stars are observed centered at the Sun. The portal ingests meteor data through the backend to extract insights and visualize them. The easily digestible overview of the data allows space scientists to maneuver the data, assisting scientific discovery and data exploration. The addition of an ``export timeline'' feature will enable scientists to highlight and export important trends aiding effective communication of ideas and results to a diverse audience.

This website displays meteor observations that took place on a particular date and at a specific location. Meteors clusters are colored (blue, green, yellow, orange, and red; velocity lower to higher) based on their velocities, while sporadic meteors are white in color. Meteor-specific information such as velocity, solar longitude, ecliptic latitude, and ecliptic longitude can be seen by clicking on every meteor. Space view is also available for every meteor. To track a meteor shower, users can set the video option, which provides a video of the meteor shower path \cite{githubGitHubSidgancamspapervids}. The portal includes constellations as landmarks to aid in location identification. Constellation coordinates are programmatically generated to reflect the unique movement of each star while accommodating potential shear. Users can also hover over constellations to highlight the meteors which originate there and the direction they fall towards the earth.  The portal also allows the ability to zoom in or out of a meteor shower, allowing scientists to view showers hidden or overlaid by others. Responsive latitude and longitude coordinates move as the celestial sphere is rotated or zoomed into. This portal also provides the an interactive space view of meteors.  Using space view API \cite{ianwwWebster} this feature has been integrated with CAMS Meteor Shower Portal. For any given International Astronomical Union (IAU) number of a meteor, we can see the solar system view, follow earth watch view, from earth view of that meteor. The source code of this portal is publicly available \cite{githubGitHubSidgannasaweb}. This web portal is officially launched on 10th November, 2022 during a SETI Live event \cite{youtubeSETILive, youtubeCAMS}.

\section{Ablation Analysis}
\label{sec:ablation_analysis}

\begin{figure*}[ht]
  \begin{subfigure}[b]{0.5\linewidth}
    \centering
    \includegraphics[width=\linewidth, height = 4cm]{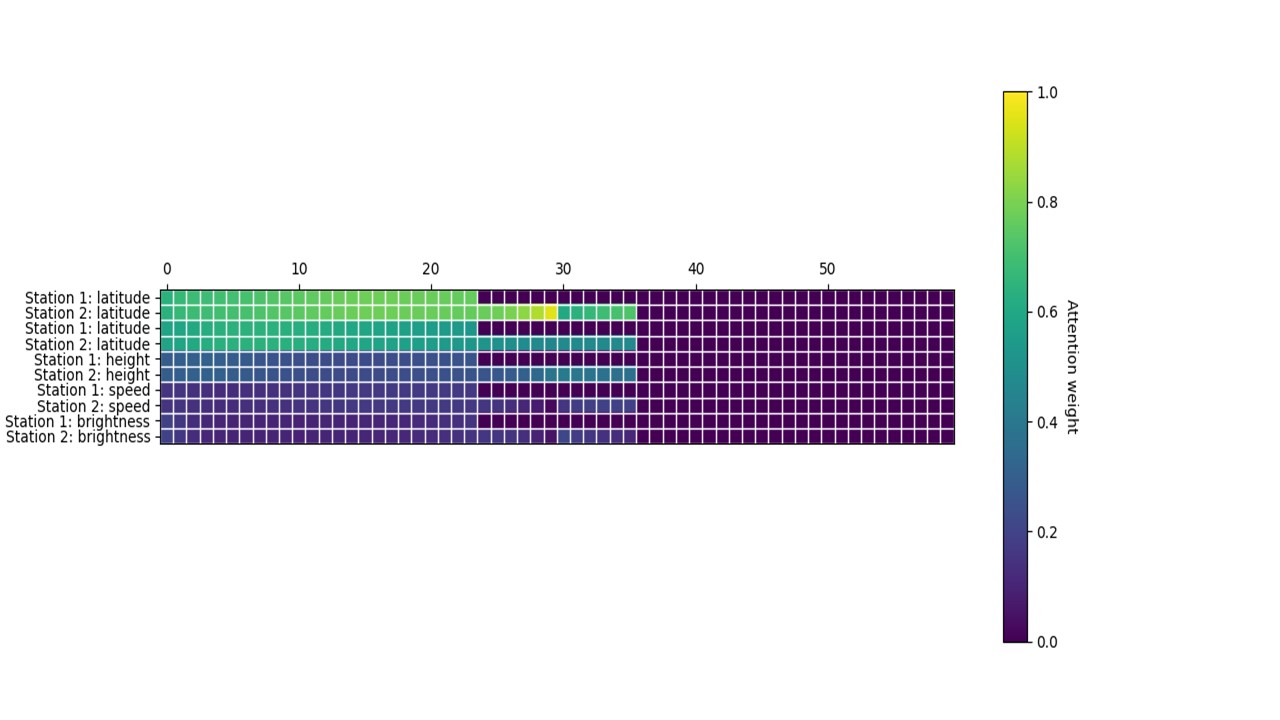} 
    \caption{Heatmap of BiLSTM weights for positive example} 
    \label{fig:postive_heatmap} 
    \vspace{4ex}
  \end{subfigure}
  \begin{subfigure}[b]{0.5\linewidth}
    \centering
    \includegraphics[width=\linewidth, height = 4cm]{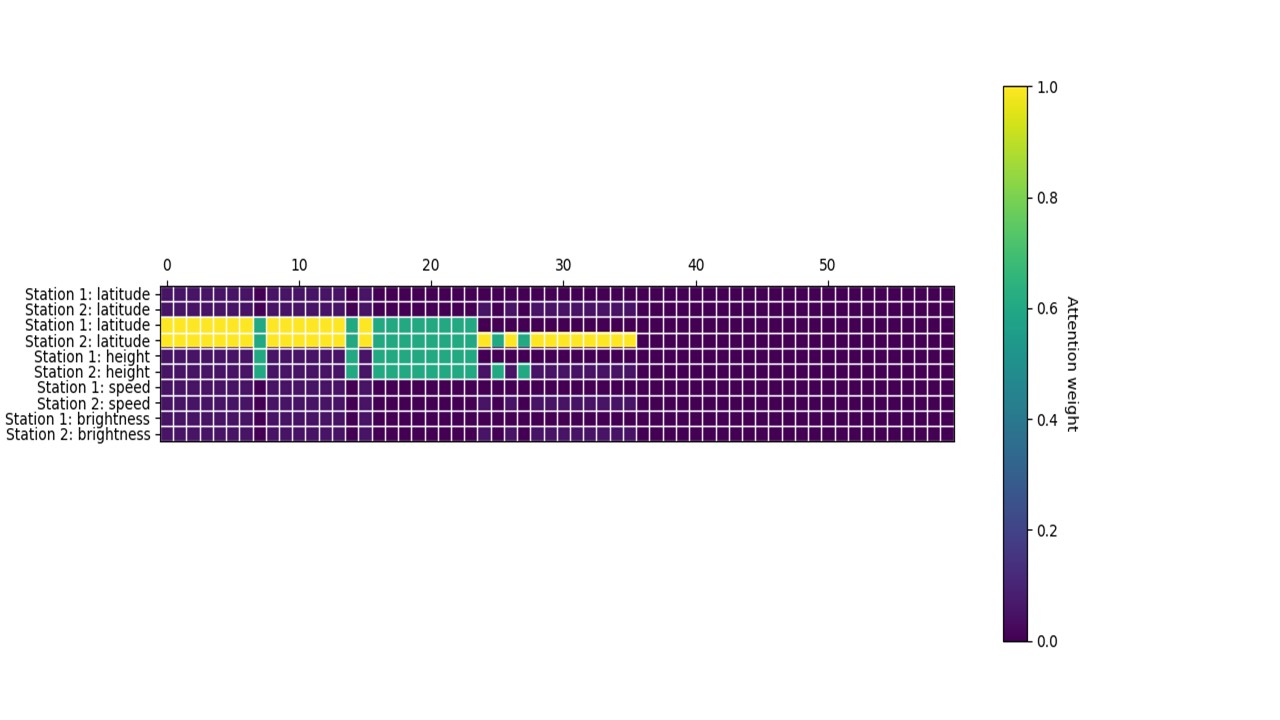} 
    \caption{Heatmap of BiLSTM weights for negative example} 
    \label{fig:negative_heatmap} 
    \vspace{4ex}
  \end{subfigure} 
  \begin{subfigure}[b]{0.5\linewidth}
    \centering
    \includegraphics[width=0.9\linewidth, height = 4cm]{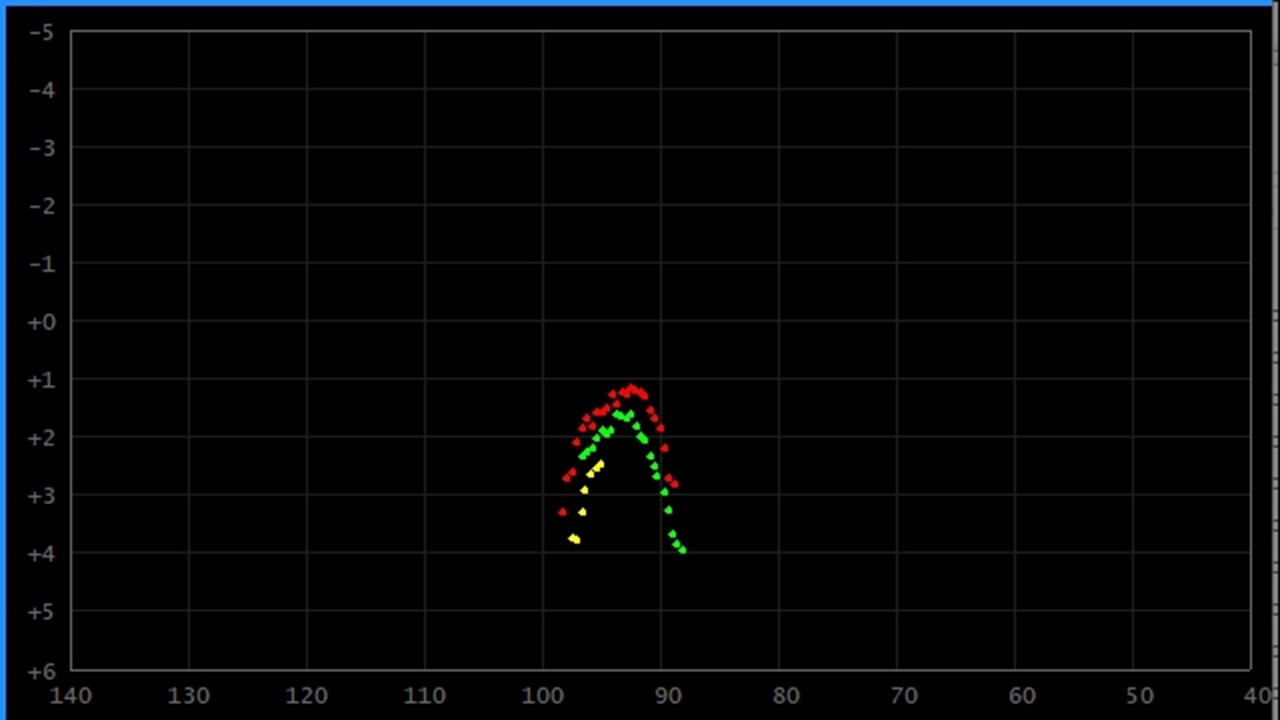} 
    \caption{Light curve of positive example} 
    \label{fig:postive_light} 
  \end{subfigure}
  \begin{subfigure}[b]{0.5\linewidth}
    \centering
    \includegraphics[width=0.9\linewidth, height = 4cm]{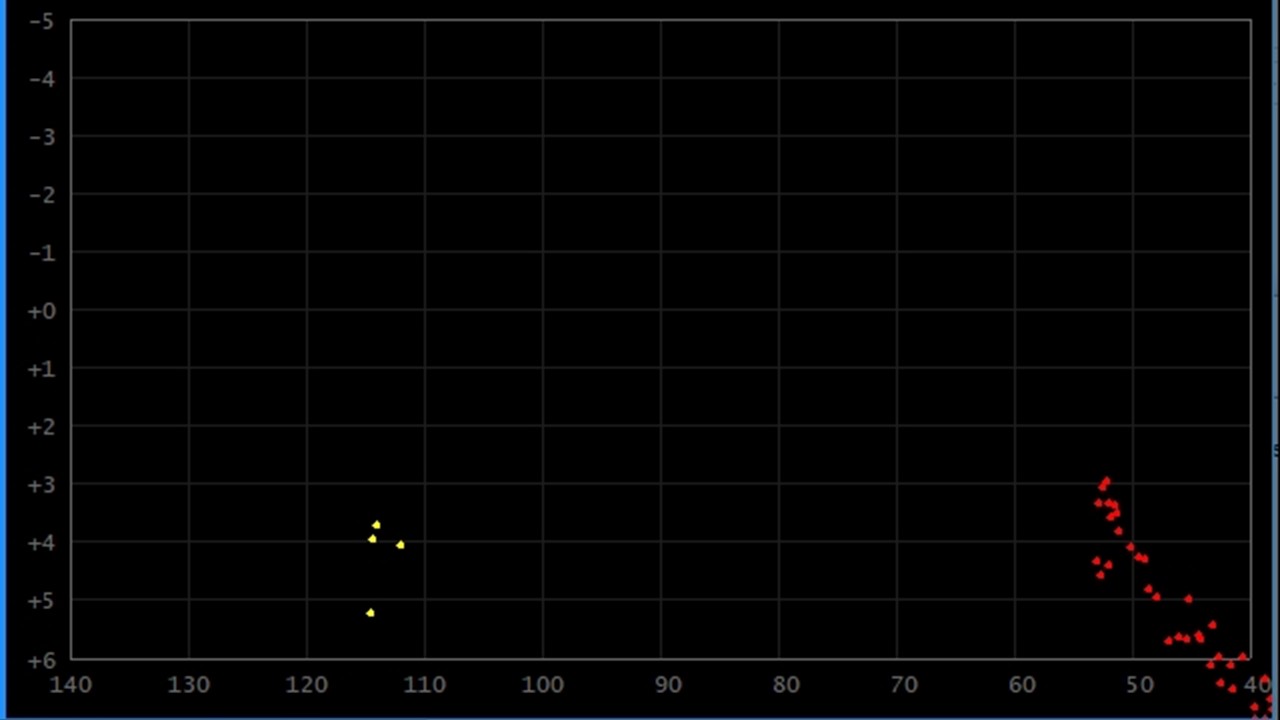} 
    \caption{Light curve of positive example} 
    \label{fig:negative_light} 
  \end{subfigure} 
  \caption{Illustration ablation analysis}
  \label{fig:analysis} 
\end{figure*}

In this section, we are making an attempt to explain how the classifier predicts the class. To understand the classifier we use a deep attention-based LSTM model and we capture the weights of the attention layer and plot it. In Fig \ref{fig:postive_heatmap} and Fig \ref{fig:negative_heatmap} we can find a heatmap of weights for a positive and a negative example. In the positive example, the weight changes gradually and maintains a continuous pattern. On the other hand, in the negative example, the weight change is discrete and maintains no such pattern. This pattern shows a resemblance with the light curves of the positive and negative examples in Fig \ref{fig:postive_light} and Fig \ref{fig:negative_light}. This gives us confidence in the decision-making capability of the classifier and also helps us verify that the AI model is attempting to replicate a scientist's methodology.

Fig \ref{fig:postive_heatmap} and Fig \ref{fig:negative_heatmap} are the heatmap of the attention weights of five features for the positive and negative examples of the meteor shower. The figures represent meteor showers observed from two stations (Station 1 and Station 2). The figures show the data for 60 frames, the first 24 frames are observed from Station 1 and the rest are observed from Station 2. In Fig \ref{fig:postive_heatmap}, we can observe a continuous distribution of the weights, while in Fig \ref{fig:negative_heatmap} the weights are discrete.

\section{Results}
\label{sec:results}

\subsection{Global Scaling of CAMS}
\label{sec:gloabal_scaling}

Data pipeline automation and ML algorithm refinement have led to a rapid expansion of the CAMS Network. Over the past five years, CAMS network coverage has increased six-fold (600+ cameras today). This expansion includes first light in the Southern Hemisphere and Asia, enabling 24x7 global meteor coverage. The increase in camera numbers has produced larger volumes of data and better statistical estimation. Apart from the improved efficiency offered by automation as a part of network extension, the CAMS network expansion has raised awareness and encouraged citizen science participation through the use of inexpensive equipment to improve the understanding of meteor showers.
The global CAMS network continued to operate well during 2022 and increased its combined annual yield to about 510,000 orbits/year, up from last year's 470,000. The individual preliminary tallies, with 2021 and 2022 results, are provided in Table \ref{tab:tally}. The yield increases for CAMS California and the United Arab Emirates Astronomical Camera Network are the result of an upgrade to new Watec Wat902 H2 Ultimate cameras in 2021. This doubled the yield for the UACN. 

The new CAMS India network is slowly being assembled and hopefully will achieve first light in the new year. Much of the remaining growth this year came as a result of the addition of RMS cameras to the LO-CAMS, BeNeLux, and Australia networks. These cameras were also reported to the Global Meteor Network (Denis Vida). David Rollinson has worked out a way to submit past RMS camera data, as a result of which the 2022 tally for CAMS Australia is expected to increase when that effort is completed. Arizona's LO-CAMS is now the most prolific network in the Northern Hemisphere, while CAMS Namibia keeps that crown in the Southern Hemisphere. 

\subsection{Important Findings}
\label{sec:important_findinfs}

The global CAMS has been functioning efficiently since 2011 and the size of the network is increasing every year. The CAMS network discovers new meteor showers, validates previously reported showers, demonstrates the presence of yet-to-be-discovered long-period comets, and improves knowledge of their orbits, which is presented in Table \ref{tab:new_meteors}.

\begin{table}[h]
\centering
\begin{tabular}{|l|c|c|}
\hline
Network & 2022    & 2021   \\ \hline
LO-CAMS  & 106,596 & 76,232 \\ \hline
Namibia & 81,197  & 99,659 \\ \hline
BeNeLux & 61,619  & 47,023 \\ \hline
California      & 49,051  & 51,350 \\ \hline
Australia     & 38,114  & 54,893 \\ \hline
United Arab Emirates     & 32,597  & 16,294 \\ \hline
Florida      & 26,454  & 24,554 \\ \hline
New Zealand      & 16,856  & 21,661 \\ \hline
Arizona      & 18,972  & 15,868 \\ \hline
Texas   & 19,063  & 17,449 \\ \hline
South Africa      & 7,867   & 8,726  \\ \hline
Maryland      & 2,384   & 5,140  \\ \hline
Turkey  & 1,605   & 1,323  \\ \hline
Brazil  & 105     & 144    \\ \hline
\end{tabular}
\caption{Number of observations by CAMS in 2021 and 2022}
\label{tab:tally}
\end{table}

\begin{table*}[h]
\centering
\begin{tabular}{|l|c|c|c|}
\hline
Meteor Shower Name                                                           & IAU Code & IAU  & Year \\ \hline
tau-Herculids \cite{meteornewsAnticipatingMeteor}           & TAH      & 0061 & 2022 \\ \hline
August delta-Capricornids   \cite{jenniskens2022august}     &     ADC     &   199   & 2022 \\ \hline
Arids \cite{meteornewsFirstDetection}                       &     ARD    & 1130 & 2021 \\ \hline
June theta2-Sagittariids \cite{juneTheta2Sagittariids}      &   JTT      & 1129 & 2021 \\ \hline
gamma Crucids \cite{meteornewsOutburstGamma}                & GCR      & 1047 & 2020 \\ \hline
29 Piscids \cite{setiSEENBEFORE}                            & PIS      & 1046 & 2020 \\ \hline
September upsilon Taurids \cite{meteornewsSeptemberUpsilon} & SUT      & 1045 & 2020 \\ \hline
gamma-Piscis Austrinids  \cite{meteornewsOutburstGamma}       &    GPA     & 1036 & 2020 \\ \hline
sigma Phoenicids \cite{nvidiaCAMSSystem}                    & SPH      & 1035 & 2020 \\ \hline
chi Cygnids \cite{jenniskens2015new}                        & CCY      & 757  & 2015 \\ \hline
Volantids \cite{jenniskens2016surprise}                     & VOL      & 758  & 2015 \\ \hline
\end{tabular}
\caption{Newly detected showers with unusual meteor activity}
\label{tab:new_meteors}
\end{table*}

The CAMS network contributed to the discovery of  12 new meteor showers including Arids (IAU $\#$1130), June theta2 Sagittariids (IAU $\#$1129), gamma Crucids (IAU $\#$1047), 29 Piscids (IAU $\#$1046), September upsilon Taurids (IAU $\#$1045), gamma Piscis Austrinids (IAU $\#$1036), sigma Phoenicids (IAU $\#$1035), chi Cygnids (IAU $\#$757), Volantids (IAU $\#$758),  tau-Herculids, and August delta-Capricornids. The details are presented in Table [4]. On 13th-14th December 2017, the CAMS network also detected  the highest number of meteors in a single night (including 3003 Geminids $\&$ 1154 sporadic meteors) in NASA's 63-year history. On May 31 and  August 17 of 2022, the network directed tau-Herculids outburst and an unexpected outburst of August delta-Capricornid. August delta-Capricornid is the first recorded meteor activity from comet 45P/Honda-Mrkos-Pajdusakova. We also reported on the search for meteors from the asteroid Bennu. 

Moreover, CAMS also produced the first-ever instrumental evidence of the Grigg-Mellish comet, as well as observations of rare showers including A-Carinid, chi Phoenicids, Ursids, delta Mensid, rho Phoenicids, chi Cygnids, 15 Bootids, June epsilon Ophiuchids, Phoenicids, alpha Monocerotid, Draconids, and October Camelopardalids.

\section{Acknowledgement}
We want to extend our heartfelt thanks to Dr. Burcin Bozkaya for providing invaluable guidance and support throughout our research efforts. We would also like to acknowledge the contributions of Dimitri Angelov, Steven Spielman, Amanda Norton, Timothy McCormack,  Vivienne Prince, Reilly Kalani Stanton, and Sara Haman from New College of Florida, who helped develop our ML and DL models and experiments. Lastly, we are grateful to Alfred Emmanuel, Chad Roffey, Jesse Lash, Julia Nguyen, and Meher Anand Kasam for their diligent work in creating the meteor shower portal.

\section{Appendix A}
\label{appendix_a}
This section provides a detailed explanation of all the experiments. We employed  four distinct deep learning models: LSTM, CNN-LSTM, GRU, and CNN-GRU, to create 18 models. Based on the data usage, these models can be classified into two groups: serial and parallel. 
Out of the 18 models, 14 are serial models (Baseline, LSTM-1, LSTM-2, LSTM-3, GRU-1, GRU-2, GRU-3, CNN-LSTM-1, CNN-LSTM-2, CNN-LSTM-3, CNN-GRU-1, CNN-GRU-2, CNN-GRU-3, LeakyPostReg) while the remaining four 
(P-CNN-GRU-3, P-BiLSTM, P-CNN-BiLSTM-1, P-CNN-BiLSTM-2) are parallel models. Details of each model are provided in the subsequent sections.

\section{Model Architectures}
\label{model_architecture}

\subsection{LeakyPostRegLSTM}
This model \ref{model:leakypostreg} has the same structure and dimensions as the baseline models \ref{model:baseline}, but with batch normalization incorporated. The activation function used in the dense layers is LeakyReLU, which is different from ReLU. LeakyReLU multiplies values less than 0 by a scalar value (alpha) set by the engineer. In our tests, we set alpha to 0.1, meaning negative activation values were 1/10th of their original value. In this model, the batch normalization is applied before the activation function between the dense layers.

\subsection{LSTM Models}
We have created three different versions of the baseline model by incorporating varying numbers of BiLSTM layers (1, 2, and 3) \ref{model:lstm3}. Each model is named based on the number of BiLSTM layers it contains. Additionally, these models differ from the base model in their use of LeakyReLU as the activation function.

\subsection{GRU Models}
\begin{figure*}[ht]
    \centering
    \begin{subfigure}[t]{0.45\textwidth}
        \centering
        \includegraphics[height=2.5in, width = 1.5in]{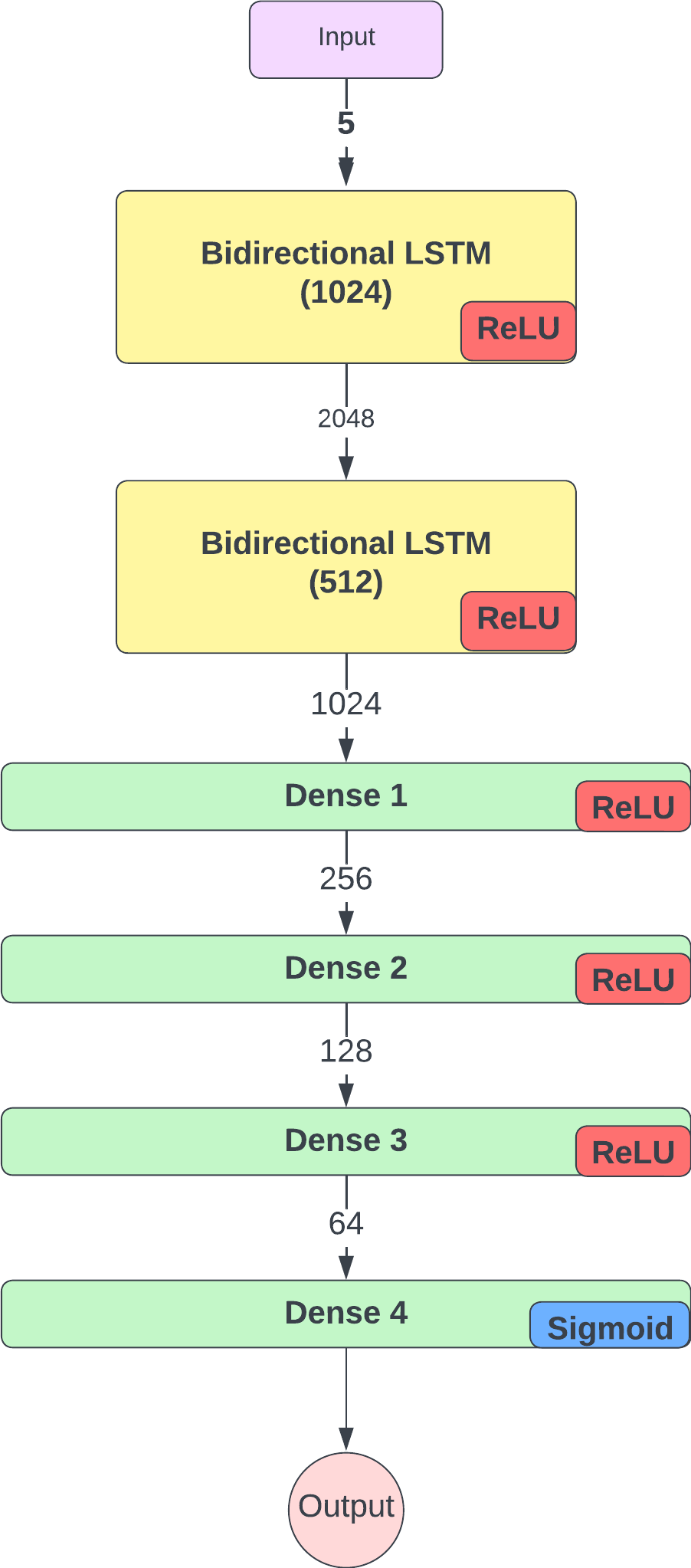}
        \caption{P-BiLSTM}
        \label{model:baseline}
    \end{subfigure}%
    ~ 
    \begin{subfigure}[t]{0.45\textwidth}
        \centering
        \includegraphics[height=2.5in, width = 1.5in]{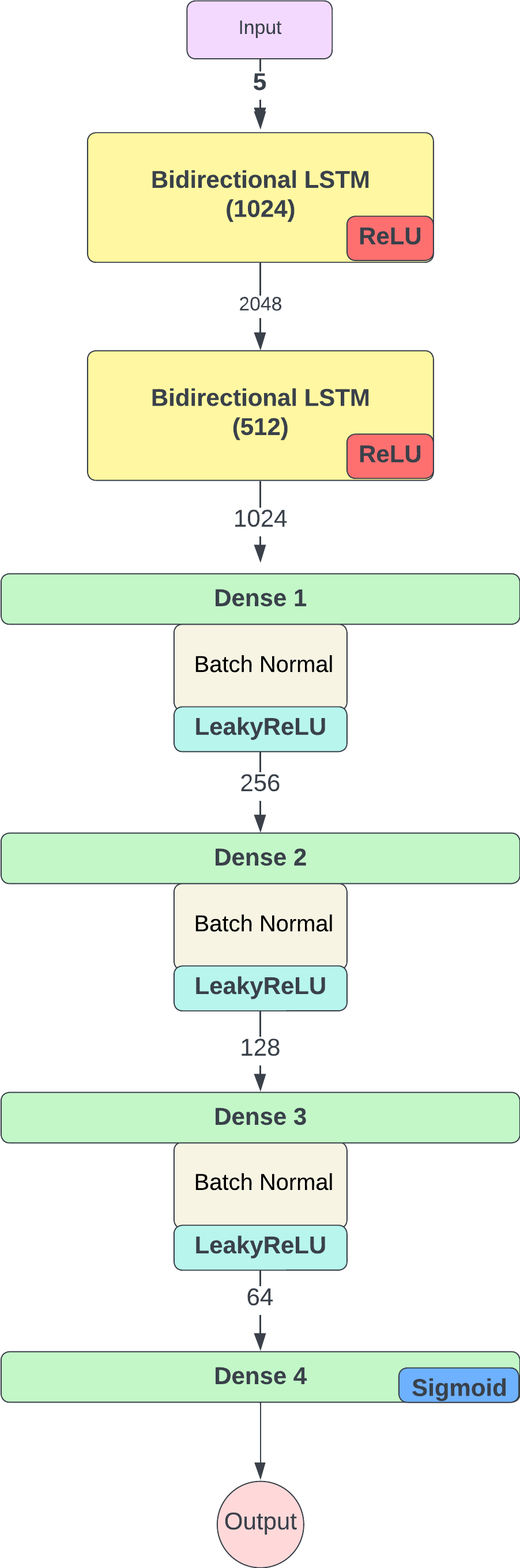}
        \caption{P-CNN-GRU-3}
        \label{model:leakypostreg}
    \end{subfigure}
    \medskip
    \begin{subfigure}[t]{0.45\textwidth}
        \centering
        \includegraphics[height=2.5in, width = 1.5in]{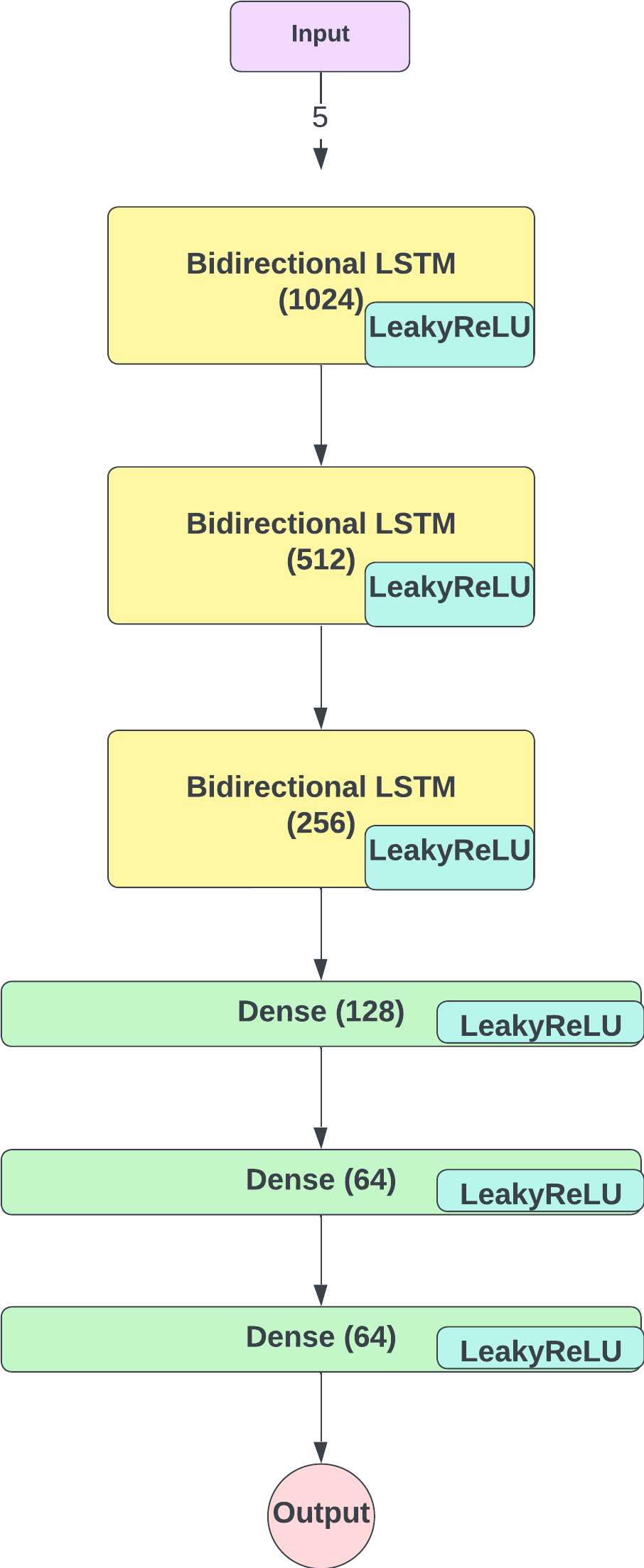}
        \caption{P-CNN-BiLSTM-1}
        \label{model:lstm3}
    \end{subfigure}
    ~
    \begin{subfigure}[t]{0.45\textwidth}
        \centering
        \includegraphics[height=2.5in, width = 1.5in]{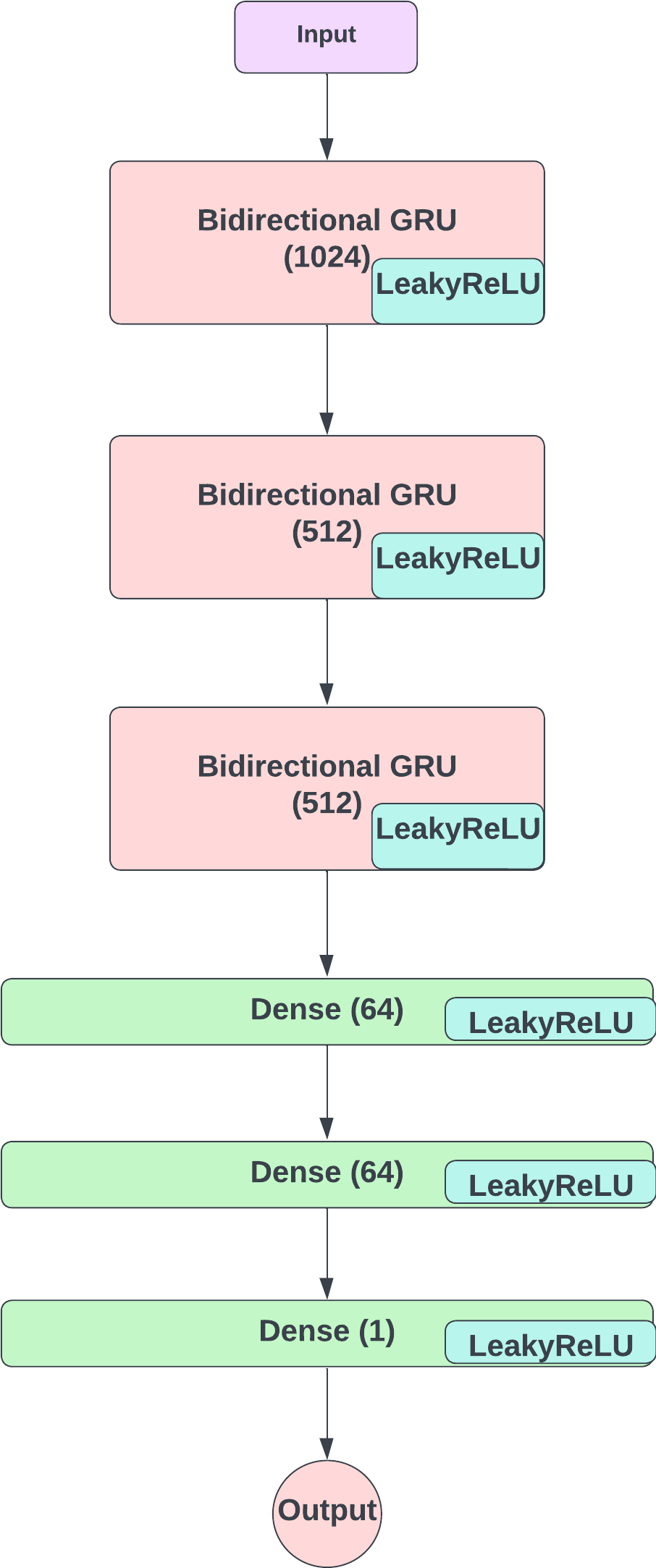}
        \caption{P-CNN-BiLSTM-2}
        \label{model:gru3}
    \end{subfigure}
    ~
    \begin{subfigure}[t]{0.45\textwidth}
        \centering
        \includegraphics[height=2.5in, width = 1.5in]{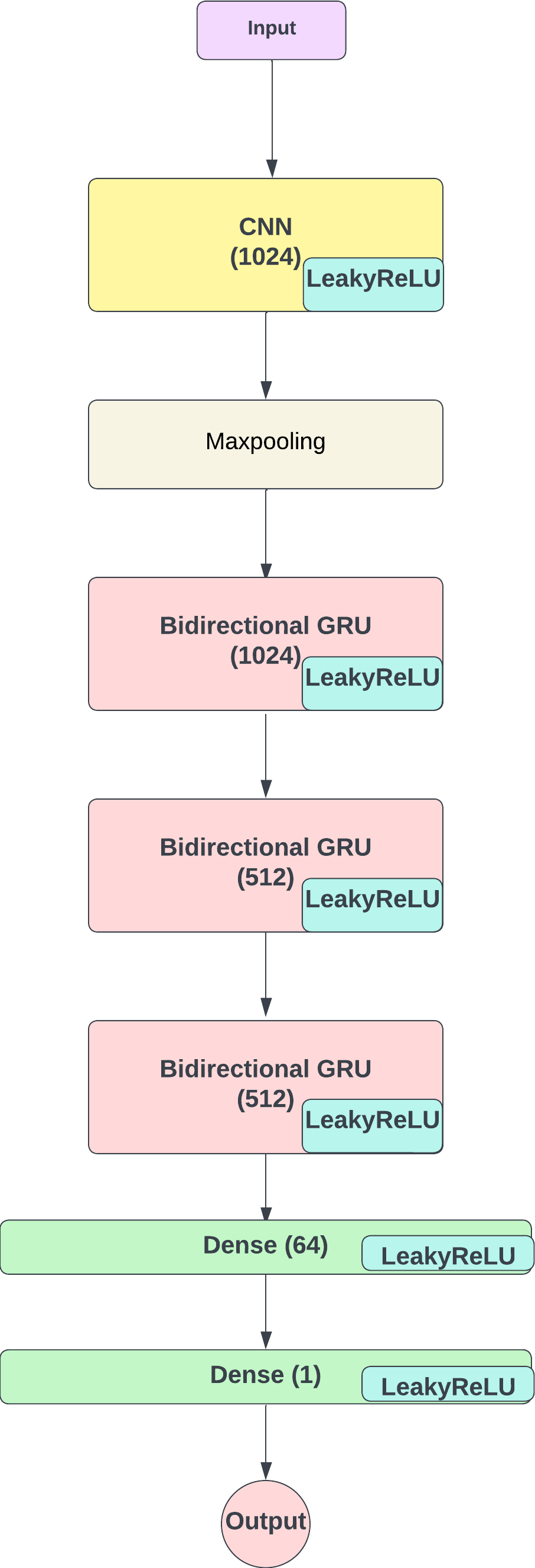}
        \caption{P-CNN-BiLSTM-1}
        \label{model:cnngru3}
    \end{subfigure}
    ~
    \begin{subfigure}[t]{0.45\textwidth}
        \centering
        \includegraphics[height=2in, width = 1.5in]{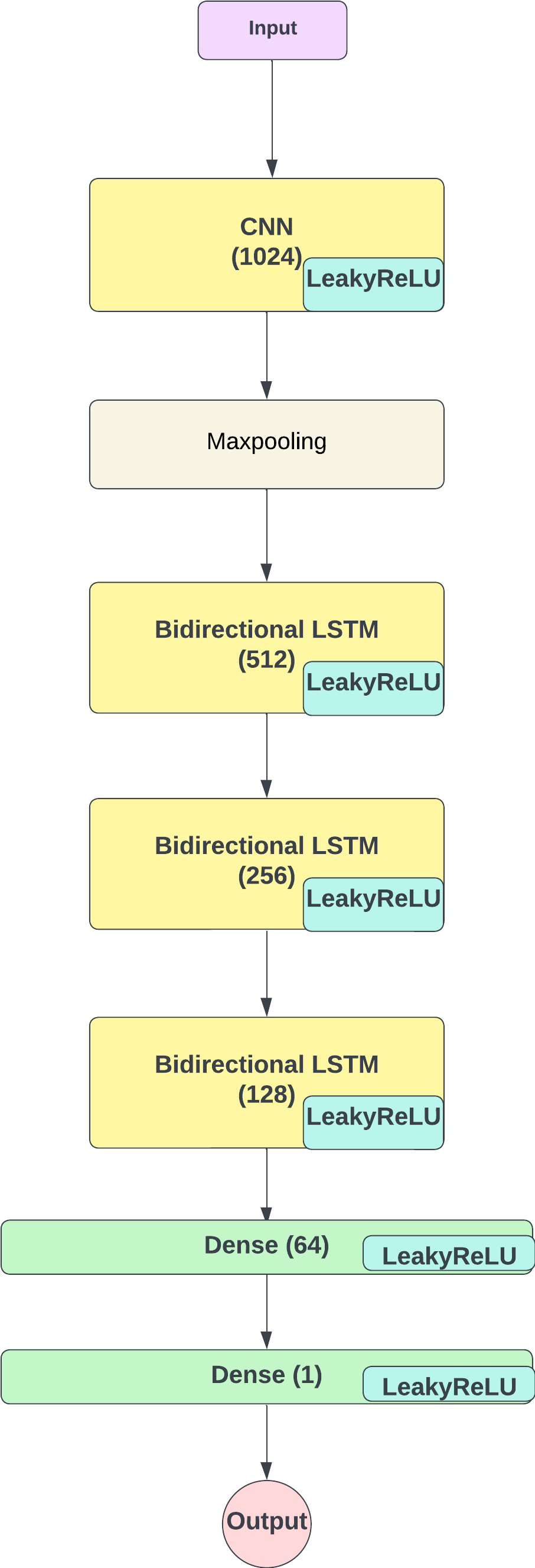}
        \caption{P-CNN-BiLSTM-2}
        \label{model:cnnbilstm3}
    \end{subfigure}
    \caption{Parallel Architecture}
\end{figure*}

We conducted experiments on three different variants of GRU models, similar to the LSTM models. These models have different numbers of GRU layers (1, 2, and 3) \ref{model:gru3} and utilize the LeakyReLU activation function. Similar to the LSTM models, we have experimented with three different variants of GRU models by setting different numbers (1, 2, and 3) of GRU layers. These models are also using the LeakyReLU activation function.

\subsection{CNN Hybrid Architecture}

\begin{figure*}[ht]
    \centering
    \begin{subfigure}[t]{0.45\textwidth}
        \centering
        \includegraphics[height=2.5in, width = 1.5in]{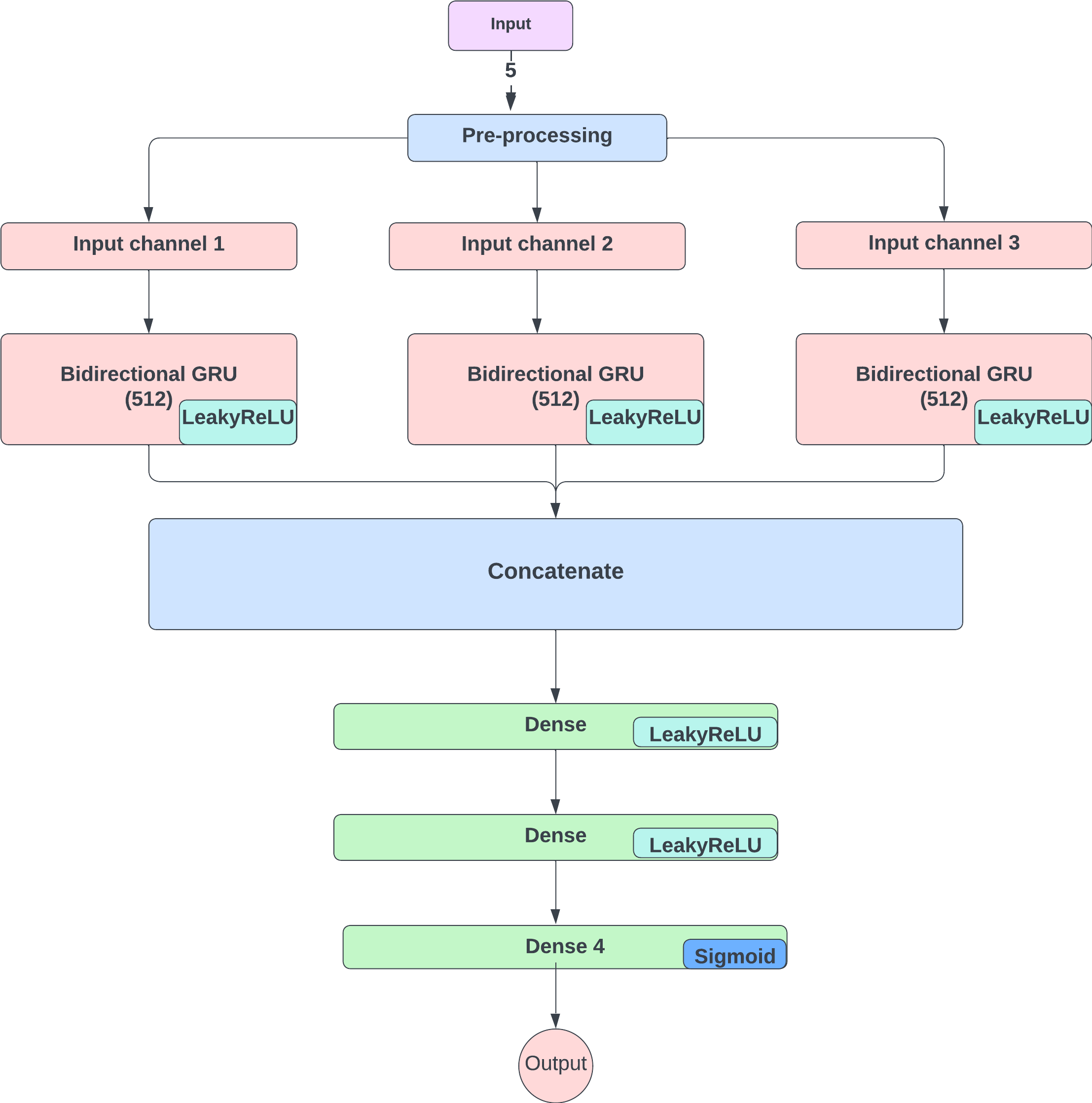}
        \caption{P-BiLSTM}
        \label{model:pbiLSTM}
    \end{subfigure}%
    ~ 
    \begin{subfigure}[t]{0.45\textwidth}
        \centering
        \includegraphics[height=2.5in, width = 1.5in]{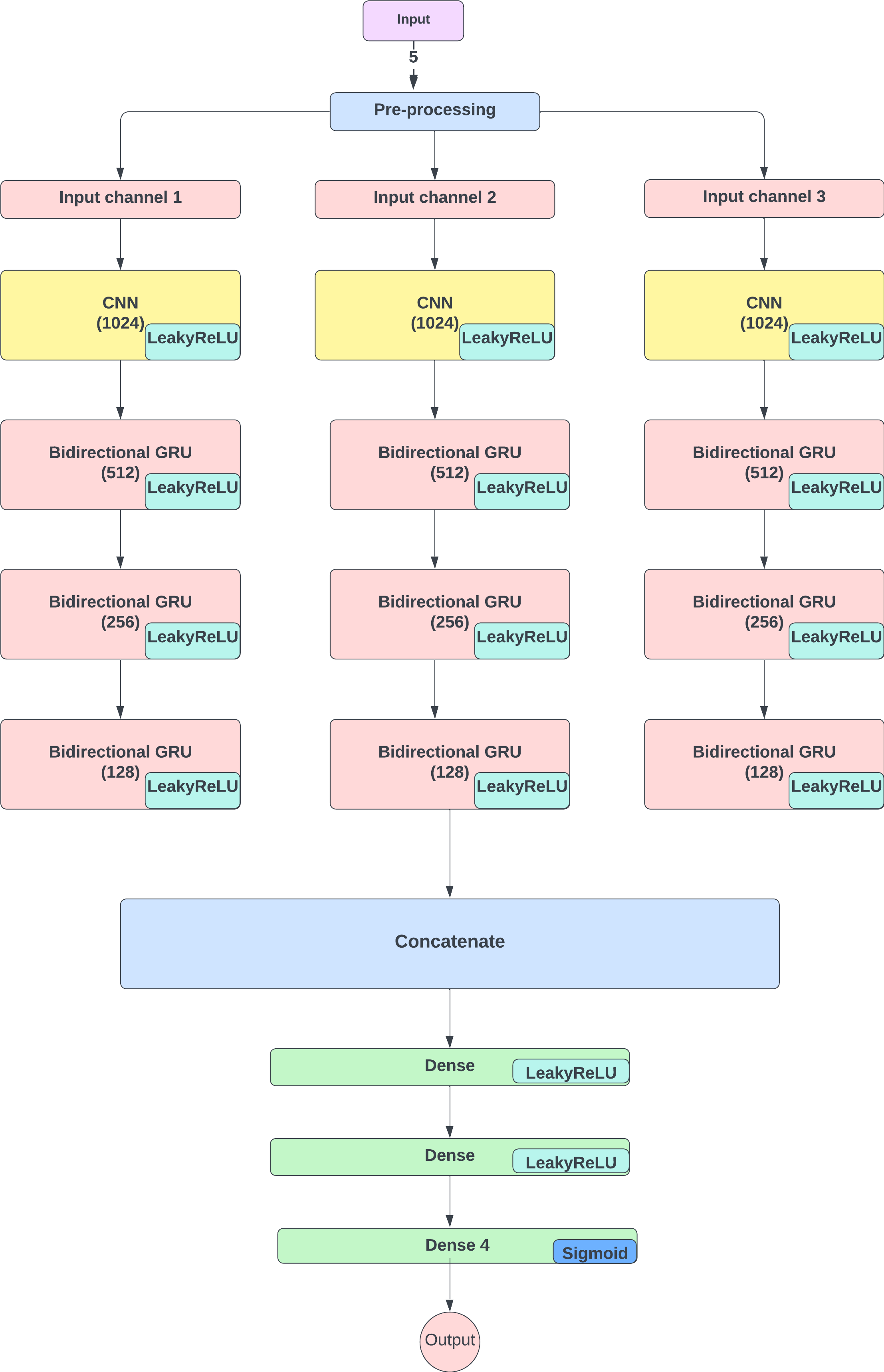}
        \caption{P-CNN-GRU-3}
        \label{model:pcnngru3}
    \end{subfigure}
    \medskip
    \begin{subfigure}[t]{0.45\textwidth}
        \centering
        \includegraphics[height=2.5in, width = 1.5in]{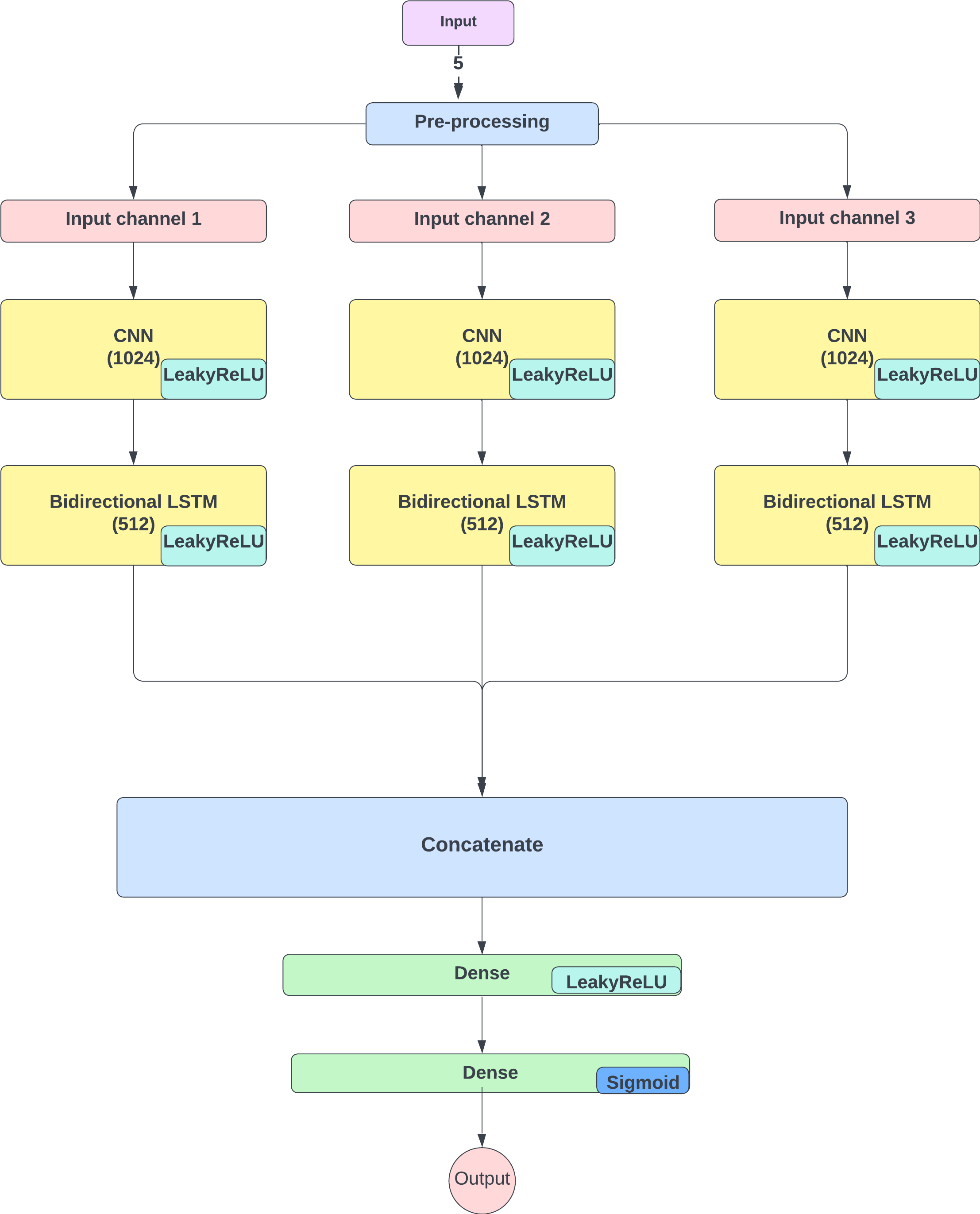}
        \caption{P-CNN-BiLSTM-1}
        \label{model:pcnnbilstm1}
    \end{subfigure}
    ~
    \begin{subfigure}[t]{0.45\textwidth}
        \centering
        \includegraphics[height=2.5in, width = 1.5in]{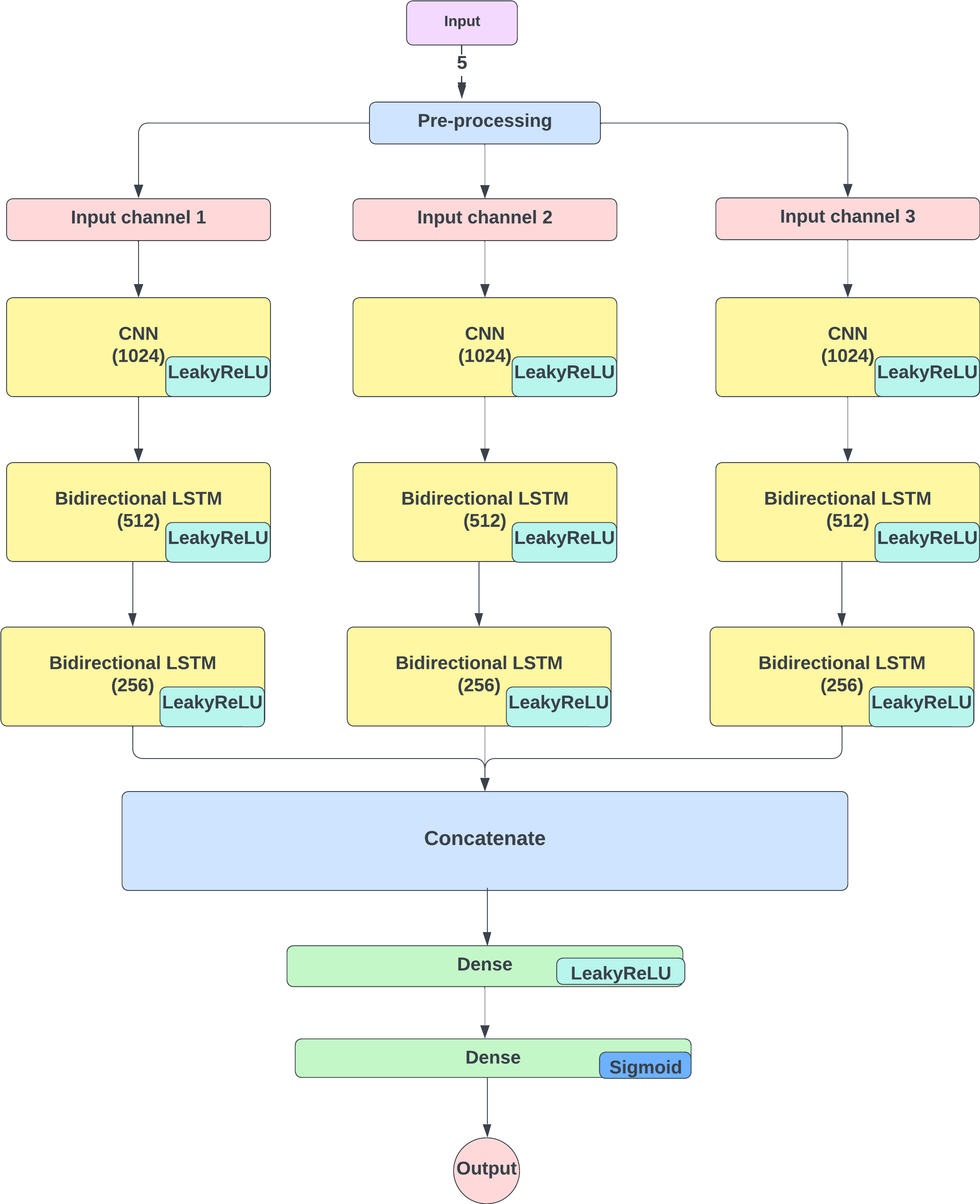}
        \caption{P-CNN-BiLSTM-2}
        \label{model:pcnnbilstm2}
    \end{subfigure}
    \caption{Parallel Architecture}
\end{figure*}

CNN-based hybrid models \ref{model:cnngru3} and \ref{model:cnnbilstm3} include a convolutional pooling layer to extract features from the meteor sequences. To leverage the convolutions as feature extraction of CNN, the model can pick up additional spatial context within the sequences to increase performance. CNNs excel at learning spatial structure in input data and the  CAMS data have a dimension spatial structure in the sequence of numeric information in the textfiles, so the CNN may be able to pick out invariant features for signs of which features match a meteor and which features aren’t a meteor. These learned spatial features from the variables we feed into our model can then be learned as sequences by an LSTM layer before it is fed into dense layers to get our single output.

The CNN-GRU architecture has a similar motivation as the CNN-LSTM
architecture due to the hybrid nature of both architectures the hybrid models have the convolutional layers preceding the recurrent layers (either LSTM or GRU). There is less precedent for CNN-GRU models than for CNN-LSTM models. Since GRU architectures have fewer parameters than LSTM architectures, the GRU models have faster runtimes.

\subsection{Parallel Models}
We developed four parallel models: a parallel BiLSTM (P-BiLSTM) \ref{model:pbiLSTM}, a parallel stacked CNN-BiLSTM (P-CNN-BiLSTM-1) \ref{model:pcnnbilstm1}, a parallel stacked CNN-BiLSTM with two Bi-LSTMS (P-CNN-BiLSTM-2) \ref{model:pcnnbilstm2} and a parallel stacked CNN-GRU with three GRU layers (P-CNN-GRU-3) \ref{model:pcnngru3}. The models will be referred to by these acronyms for the rest of this paper. The parallel BiLSTM  (P-BiLSTM) differs from the model implemented in \cite{bouaziz2016parallel}  by incorporating bidirectional LSTMs rather than simply parallel LSTMs that feedforward separate sequences.

\begin{table}[H]
\begin{tabular}{|l|l|c|}
\hline
\multicolumn{1}{|c|}{Model Name} & \multicolumn{1}{c|}{\begin{tabular}[c]{@{}c@{}}Parallel Block \\ (3 channels)\end{tabular}}   & \begin{tabular}[c]{@{}c@{}} Dense \\ Layers \end{tabular} \\ \hline
P-BiLSTM                         & 1 BiLSTM Layer                                                                   & 4                                                                 \\ \hline
P-CNN-BiLSTM-1                   & \begin{tabular}[c]{@{}l@{}}1 Convolutional layer, \\ 1 BiLSTM Layer\end{tabular} & 2                                                                 \\ \hline
P-CNN-BiLSTM-2                   & \begin{tabular}[c]{@{}l@{}}1 Convolutional layer, \\ 2 BiLSTM Layer\end{tabular} & 2                                                                 \\ \hline
P-CNN-GRU-3                      & \begin{tabular}[c]{@{}l@{}}1 Convolutional layer, \\ 3 BiLSTM Layer\end{tabular} & 3                                                                 \\ \hline
\end{tabular}
\caption{Description of each parallel block and the number of dense layers}
\label{parallel_block}
\end{table}

The P-BiLSTM \ref{model:pbiLSTM} is an approximate implementation of the LeakyPostReg BiLSTM (the best baseline alteration) in a parallel architecture. The difference is that the LeakyPostReg BiLSTM has two stacked BiLSTMS whereas the P-BiLSTM only has a single BiLSTM. The decision to use only a single BiLSTM was because of the exploding training time when stacked BiLSTMs were used ($>$18 hours on a single GPU). 

To maximize comparability, the four models are structurally similar; the type of the recurrent  layers is switched between the models but all else (the objective function, activation functions, etc.) are held constant. Because the LSTMs stacked with CNNs become very complex, more complex than these data may require, we found that reducing the number of dense layers from 4 to just 2 decreased over-fitting and led to a non-trivial reduction in training time. 

Another difference from the baseline model is that we do not scale the input data for the parallel models. Both the MinMax scaler, which is used in the baseline code, and the StandardScaler (the other common scaler) are sensitive to outliers. We identified outliers in several features, which indicates that these scalers are unsuitable. Further, while tuning the models, we found that they performed no better when the input data was scaled compared to when it was not, so we removed the scaler.

Batch normalization is used in between the dense layers in each model, except before the last one \cite{ioffe2015batch}. Recent research has cautioned against using combinations of batch normalization and dropout; and theoretically, batch normalization annuls the need for dropout, so we did not add further regularization to any of the models. The weights for the dense layers are initialized using He initialization (also known as Kaiming initialization) \cite{he2015delving}. Drawing off of our results from the baseline alterations, all of the layers use the LeakyReLU activation function. The maximum length for each sequence is 45. This is a tunable parameter. Because the best performing baseline models use a segment length of 45, we use a segment length of 45 for all of the parallel models to improve comparability.

Each model is trained within a designated script for that model. The training procedure for each model is identical. Identical to the baseline models, the parallel models were built using tensorflow/keras  and were trained with consistent dimensions. All four use the Adaptive Moments optimizer (Adam) with an initial learning rate between 0.00001 and 0.0000025, binary cross entropy loss function, and a batch size of 100 trained over 50 epochs max. The initial learning rate was tuned for each model using the validation set. Additionally, all models are provided with the call backs for early stopping and restoring best weights (via lowest validation loss). Models return logits from a sigmoid activation function that corresponds to the input sequence.

{\small
\bibliographystyle{ieee_fullname}
\bibliography{PaperForReview}

\begin{thebibliography}{10}\itemsep=-1pt

\bibitem{githubGitHubSidgannasaweb}
{G}it{H}ub - sidgan/nasaweb: {N}{A}{S}{A} {C}{A}{M}{S} ({C}amera for {A}ll
  {S}ky {S}urveillance) {M}eteor {S}hower {P}ortal by {S}pace{M}{L} ---
  github.com.
\newblock \url{https://github.com/sidgan/nasaweb}, 2023.
\newblock [Accessed 21-Jun-2023].

\bibitem{setiNASACAMS}
{N}{A}{S}{A} {C}{A}{M}{S} {M}eteor {S}hower {P}ortal ---
  meteorshowers.seti.org.
\newblock \url{https://meteorshowers.seti.org/}, 2023.
\newblock [Accessed 21-Jun-2023].

\bibitem{juneTheta2Sagittariids}
SETI~Institute Astronomy.
\newblock {N}{E}{W} {S}{H}{O}{W}{E}{R} {D}{E}{T}{E}{C}{T}{E}{D}: {J}{U}{N}{E}
  {T}{H}{E}{T}{A}2 {S}{A}{G}{I}{T}{T}{A}{R}{I}{I}{D}{S} --- seti.org.
\newblock
  \url{https://www.seti.org/new-shower-detected-june-theta2-sagittariids},
  2021.
\newblock [Accessed 23-Jun-2023].

\bibitem{bland2012australian}
PA Bland, P Spurn{\`y}, AWR Bevan, KT Howard, MC Towner, GK Benedix, RC
  Greenwood, L Shrben{\`y}, IA Franchi, G Deacon, et~al.
\newblock The australian desert fireball network: A new era for planetary
  science.
\newblock {\em Australian Journal of Earth Sciences}, 59(2):177--187, 2012.

\bibitem{dfs}
Philip~A Bland.
\newblock {The Desert Fireball Network}.
\newblock {\em Astronomy \& Geophysics}, 45(5):5.20--5.23, 10 2004.

\bibitem{bouaziz2016parallel}
Mohamed Bouaziz, Mohamed Morchid, Richard Dufour, Georges Linar{\`e}s, and
  Renato De~Mori.
\newblock Parallel long short-term memory for multi-stream classification.
\newblock In {\em 2016 IEEE Spoken Language Technology Workshop (SLT)}, pages
  218--223. IEEE, 2016.

\bibitem{youtubeCAMS}
CAMS.
\newblock {C}{A}{M}{S} --- youtube.com.
\newblock
  \url{https://www.youtube.com/playlist?list=PL9qsrNYAPFKJ8lQNCco0p6q8nQ0R4O_cN}.
\newblock [Accessed 29-Jun-2023].

\bibitem{githubGitHubSidgancamspapervids}
Siddha Ganju and Amartya Hatua.
\newblock {G}it{H}ub - sidgan/cams-paper-vids --- github.com.
\newblock \url{https://github.com/sidgan/cams-paper-vids}.
\newblock [Accessed 29-Jun-2023].

\bibitem{gonzalez2018systematic}
Jos{\'e}~Nicol{\'a}s Gonz{\'a}lez~P{\'e}rez.
\newblock {\em Systematic study of the rapid optical-NIR variability of blazars
  and other AGNs}.
\newblock PhD thesis, Staats-und Universit{\"a}tsbibliothek Hamburg Carl von
  Ossietzky, 2018.

\bibitem{he2015delving}
Kaiming He, Xiangyu Zhang, Shaoqing Ren, and Jian Sun.
\newblock Delving deep into rectifiers: Surpassing human-level performance on
  imagenet classification.
\newblock In {\em Proceedings of the IEEE international conference on computer
  vision}, pages 1026--1034, 2015.

\bibitem{youtubeSETILive}
SETI Institute.
\newblock {S}{E}{T}{I} {L}ive: {I}{S} {T}{H}{A}{T} {A} {M}{E}{T}{E}{O}{R}? ---
  youtube.com.
\newblock
  \url{https://www.youtube.com/watch?v=yIu_E1U3O6c&ab_channel=SETIInstitute}.
\newblock [Accessed 21-Jun-2023].

\bibitem{setiSEENBEFORE}
SETI Institute.
\newblock {N}{O}{T} {S}{E}{E}{N} {B}{E}{F}{O}{R}{E}: {S}{A}{M}{E}
  {M}{E}{T}{E}{O}{R}{O}{I}{D} {S}{T}{R}{E}{A}{M} {S}{H}{O}{W}{S} {U}{P}
  {A}{G}{A}{I}{N} {A} {M}{O}{N}{T}{H} {L}{A}{T}{E}{R} --- seti.org.
\newblock
  \url{https://www.seti.org/not-seen-same-meteoroid-stream-shows-again-month-later},
  2020.
\newblock [Accessed 23-Jun-2023].

\bibitem{ioffe2015batch}
Sergey Ioffe and Christian Szegedy.
\newblock Batch normalization: Accelerating deep network training by reducing
  internal covariate shift.
\newblock In {\em International conference on machine learning}, pages
  448--456. pmlr, 2015.

\bibitem{jenniskens2015new}
P Jenniskens.
\newblock New chi cygnids meteor shower.
\newblock {\em Central Bureau Electronic Telegrams}, 4144:1, 2015.

\bibitem{meteornewsFirstDetection}
P Jenniskens.
\newblock {F}irst detection of the {A}rid ({A}{R}{D}, \#1130) meteor shower
  from comet 15{P}/{F}inlay --- meteornews.net.
\newblock
  \url{https://www.meteornews.net/2021/10/02/the-arid-ard-1130-meteor-shower-from-comet-15p-finlay},
  2021.
\newblock [Accessed 23-Jun-2023].

\bibitem{meteornewsOutburstGamma}
Peter Jenniskens.
\newblock {O}utburst of {G}amma {C}rucids in 2021 ({G}{C}{R}, {I}{A}{U}\#1047)
  --- meteornews.net.
\newblock
  \url{https://www.meteornews.net/2021/02/15/gamma-crucids-2021-gcr1047}, 2021.
\newblock [Accessed 23-Jun-2023].

\bibitem{meteornewsAnticipatingMeteor}
P Jenniskens.
\newblock {A}nticipating a meteor outburst: {G}lobal {C}{A}{M}{S} video network
  detects first 2022 tau {H}erculids --- meteornews.net.
\newblock
  \url{https://www.meteornews.net/2022/05/30/anticipating-a-meteor-outburst-global-cams-video-network-detects-first-2022-tau-herculids},
  2022.
\newblock [Accessed 23-Jun-2023].

\bibitem{jenniskens2022august}
P Jenniskens.
\newblock August delta capricornids meteor shower 2022.
\newblock {\em eMeteorNews}, 7(5):306, 2022.

\bibitem{jenniskens2016surprise}
Peter Jenniskens, Jack Baggaley, Ian Crumpton, Peter Aldous, Peter~S Gural,
  Dave Samuels, Jim Albers, and Rachel Soja.
\newblock A surprise southern hemisphere meteor shower on new-year's eve 2015:
  the volantids (iau\# 758, vol).
\newblock {\em JIMO}, 44:35--41, 2016.

\bibitem{nvidiaCAMSSystem}
Angie Lee.
\newblock {C}{A}{M}{S} {S}ystem {D}iscovers {N}ew {M}eteor {S}howers {U}sing
  {A}{I} | {N}{V}{I}{D}{I}{A} {B}log --- blogs.nvidia.com.
\newblock \url{https://blogs.nvidia.com/blog/2020/08/20/cams-meteor-showers},
  2020.
\newblock [Accessed 23-Jun-2023].

\bibitem{meteornewsSeptemberUpsilon}
Meteor News.
\newblock {T}he {S}eptember upsilon {T}aurid meteor shower and possible
  previous detections --- meteornews.net.
\newblock
  \url{https://www.meteornews.net/2021/01/27/the-september-upsilon-taurid-meteor-shower-and-possible-previous-detections},
  2021.
\newblock [Accessed 23-Jun-2023].

\bibitem{ianwwWebster}
Ian Webster.
\newblock {I}an {W}ebster --- ianww.com.
\newblock \url{https://www.ianww.com}.
\newblock [Accessed 29-Jun-2023].

\end{thebibliography}
}

\end{document}